\documentclass[aps,prx,twocolumn,superscriptaddress]{revtex4-2}

\usepackage[english]{babel}
\usepackage{xcolor}
\usepackage{soul}
\usepackage{enumitem}
\usepackage{multirow}
\usepackage{amsmath}
\usepackage{graphicx}
\usepackage{color}
\usepackage{combelow}
\usepackage{siunitx}
\usepackage{physics}
\usepackage{amsmath}

\begin{document}

\title{Strongly correlated physics in organic open-shell quantum systems}

\author{G.~Gandus}
 \thanks{These authors contributed equally to this work}
 %\email{gandusgui@gmail.com}
 \affiliation{Integrated Systems Laboratory, ETH Z\"urich, Gloriastrasse 35, 8092 Z\"urich, Switzerland}
\author{A.~Jayaraj}
\thanks{These authors contributed equally to this work}
 %\email{anooja.jayaraj@empa.ch}
 \affiliation{Empa, Swiss Federal Laboratories for Materials Science and Technology, {\"U}berlandstrasse 129, CH-8600, D{\"u}bendorf, Switzerland}
\author{D.~Passerone}
 \affiliation{Empa, Swiss Federal Laboratories for Materials Science and Technology, {\"U}berlandstrasse 129, CH-8600, D{\"u}bendorf, Switzerland}
\author{R.~Stadler}
 \affiliation{Institute for Theoretical Physics, Vienna University of Technology, Wiedner Hauptstrasse 8-10, A-1040 Vienna, Austria}
\author{M.~Luisier}
 \affiliation{Integrated Systems Laboratory, ETH Z\"urich, Gloriastrasse 35, 8092 Z\"urich, Switzerland}
\author{A.~Valli}
 \email{valli.angelo@ttk.bme.hu}
 %\affiliation{Institute for Theoretical Physics, Vienna University of Technology, Wiedner Hauptstrasse 8-10, A-1040 Vienna, Austria}
 \affiliation{Department of Theoretical Physics, Institute of Physics, Budapest University of Technology and Economics, M\H{u}egyetem rkp. 3., H-1111 Budapest, Hungary}
 \affiliation{HUN-REN-BME-BCE Quantum Technology Research Group, Budapest University of Technology and Economics, M\H{u}egyetem rkp. 3., H-1111 Budapest, Hungary}

\begin{abstract} 
Strongly correlated physics arises from electron-electron scattering within partially filled orbitals. 
Organic molecules in open-shell configurations are therefore good candidates to exhibit many-body effects.
%and is the case for organic molecules in an open-shell configuration.
We focus on electron transport in a two-terminal single-molecule junction setup, 
in which the molecular bridge consists of an organic radical with a molecular orbital hosting a single unpaired electron (SOMO). 
We perform beyond state-of-the-art numerical simulations combining an \textit{ab-initio} description of the chemical environment, 
with quantum field-theoretical techniques that account for many-body effects.   
The key observation is that the SOMO resonance is prone to splitting 
and we identify a \emph{giant} electronic scattering rate as the driving many-body mechanism, 
akin to that of the Mott metal-to-insulator transition. 
By comparing linear and cyclic radicals, we show that the spatial distribution of the SOMO and its projection on the molecular backbone 
have dramatic consequences for the transport properties of the junction. 
We argue that the phenomenon and the underlying microscopic mechanism apply to a broad family of open-shell molecular systems, 
and can explain puzzling experimental observations such as suppressed conductance in radical junctions. 
\end{abstract}

%needed with revtex style
\maketitle

\section{Introduction}

Strongly correlated electronic physics emerges in partially occupied orbitals in the presence of competing energy scales. The concept of a ``strongly-correlated metal" is well established, and refers to materials in which mutual Coulomb repulsion among electrons leads to collective behavior, causing the breakdown of the single-particle approximation and the emergence of complex quantum phenomena. Extending this idea to molecular systems, characterized by discrete energy levels, is not obvious. 
%
%In low-dimensional and nanoscopic systems, correlations are also expected to be enhanced by spatial confinement of the electrons.  
%in solid-state physics, the concept of a ``strongly-correlated metal" is well-established 
%for crystalline matter with a continuous distribution of states around the Fermi energy, 
%
Most stable organic molecules exhibit \textit{closed-shell} electronic configurations, 
with valence electrons of opposite spin paired such that molecular orbitals (MOs) are either empty or filled. 
%The energy difference between the frontier MOs, 
%i.e., the highest occupied (HOMO) and lowest unoccupied (LUMO) orbitals, 
%identifies the single-particle gap. 
%
%In particular, in $\pi$-conjugated systems the HOMO-LUMO gap originates from the verlap orbitals with predominant C-p$_z$ character 
%and is generally sizable: $\Delta \sim \unit{\electronvolt}$. 
%
\textit{Radical} molecules, however, are an exception to this pattern, characterized by (at least) one unpaired valence electron residing in non-bonding, singly-occupied MOs (SOMOs). Thus, radicals realize \textit{open-shell} electronic configurations, and are therefore natural candidates to exhibit strong correlation effects. 

Neutral radicals can form through bond cleavage or electron transfer 
%(e.g., in photoinduced processes) 
and often appear as intermediate products of chemical reactions. 
The presence of unpaired electrons in the outer shell makes them inherently unstable. Tremendous advances in the synthesis and characterization of radicals~\cite{chenChem7,liJACS145,shuCR123}   
have enabled the isolation of persistent radical species that, despite their high chemical reactivity, 
remain stable long enough to be trapped in the nanogap of mechanically-controlled break-junctions~\cite{bejaranoJACS140,hurtado-gallegoNL22,mitraChem2025,liuAC56,pyurbeevaNL21,naghibiAC61,pateraAC58,zhengNC11,hayakawaNL16,liuJACS135,zhangNC4,zhangJACS144,houAFM47,zhangSciAdv44,brasJPPC129} 
or deposited onto substrates~\cite{lowNL19,mishraNat598,kraneNL20,alcon2022unveiling}. 
This has facilitated the investigation of their electronic properties via transport spectroscopy. 
%This triggered a significant interest, 
Interest in such systems has grown rapidly, driven by observations of radicals exhibiting remarkable 
electronic~\cite{bejaranoJACS140,hurtado-gallegoNL22,silAC136,mitraChem2025,lawson2403.04906}, 
magnetic~\cite{herrmannJACS132,frisendaNL15,hayakawaNL16,chelliNL23}, and optical~\cite{qiuAC62,gorgonNat620,pohJACS146} functionalities. These properties are 
relevant to emerging technologies 
ranging from next-generation nano-electronics~\cite{chenChem7}, 
to spintronics~\cite{sanvitoCSR40,ishigakiCR123,herrmannJACS132,smeuPCCP23,liJACS145}, 
sensing~\cite{garnierNL20,yangACSSensors9}, and quantum information~\cite{yonekutaJACS129,zengChem7}.  
In the context of molecular electronics, noteworthy organic radicals 
include triphenylmethyl~\cite{frisendaNL15,yuanNC7,bejaranoJACS140,appeltNano10}, 
nitroxide~\cite{pyurbeevaNL21,zhangJPCC127,brasJPPC129}, 
triazine~\cite{lowNL19,jiMCF12,jiangJPCC127,mitraChem2025}, 
and benzyl~\cite{smeuJPCC114,herrmannJACS132,mitchellNC8} derivatives, 
odd-numbered polyacetylene chains~\cite{bergfieldNL11,hernangmez-prezNL20},  
and polycyclic hydrocarbons with 
non-Kekul\'e structure~\cite{zhengNC11,mishraNat598,turco2021surface,jacobPRB106,zhengNC11,kraneNL20,smeuPCCP23,turcoJACSAu3}. 
Molecular organic frameworks (such as porphine and phthalocyanine) 
hosting transition-metal centers, are typically in open-shell configuration  
and many-body effects have been shown to play a pivotal role 
in realizing tunable molecular transistors~\cite{bhandaryNA17,gaoCP6}. 

A natural question arises as to whether an organic radical retains its open-shell configuration 
in a solid-state device. 
Recent theoretical investigations of nitroxyl radicals suggest that, despite non-trivial interactions with Au electrodes,~\cite{zhangJPCC127} 
the open-shell character is robust against interactions with the solvent molecules or under applied bias~\cite{jiangJPCC127}. 
However, the experimental situation is seemingly controversial, 
as radicals appear to exhibit environment-dependent behavior~\cite{lowNL19}. 
Clear evidence of a zero-bias resonance has been reported in organic radical two-terminal junctions 
with the molecule bonding to metal electrodes~\cite{frisendaNL15,mitraChem2025,brasJPPC129} 
or adsorbed on a surface~\cite{liuJACS135,pateraAC58,zhangNC4}, 
and ascribed to the quenching of the unpaired electrons due to the Kondo effect. 
Although, in some cases, organic radicals have been shown to demonstrate enhanced conductance~\cite{liuAC56,hurtado-gallegoNL22} 
and Seebeck coefficient~\cite{hurtado-gallegoNL22} relative to their non-radical counterparts,  
the reported conductance values~\cite{liuAC56,hurtado-gallegoNL22,lowNL19} 
are typically of the order of $G \sim 10^{-3}\,G_0$. 
This is significantly lower than the expected value for a fully open transport channel exhibiting nearly resonant transmission through a low-lying molecular orbital, where $G \sim G_0 = e^2/h$.  
Moreover, conductance traces reported for a (Blatter) triazinyl radical in a break-junction setup
closely resembled those of an oxidized derivative, with conductance comparable to that of closed-shell stilbene, 
thereby suggesting that the open-shell character is lost in certain device configurations~\cite{lowNL19}.
%due to the interaction with the electrodes, the solver, or due to the applied bias~\cite{lowNL19}.  

This apparent discrepancy raises the question: is the loss of radical character the only explanation for the suppressed conductance observed experimentally? 
%or do many-body effects play a significant role?  
We argue that a possible alternative explanation can be ascribed to many-body effects. 
In organic (semiconducting) molecules and nano-structures, 
the electron-electron scattering rate is low 
due to the lack of electronic states close to the Fermi energy, 
and electronic correlations manifest in other ways. 
For instance, the failure of \emph{ab-initio} simulations in reproducing the correct HOMO-LUMO gap 
is a long-standing issue~\cite{perdewIJQC28}, 
and numerical simulations predict a many-body renormalization of the single-particle gap~\cite{senterfPRB80,hueserPRB87,valliPRB86,valliPRB91,valliPRB94,schuelerEPJST226,valliNL18,valliPRB100,pudleinerPRB99,fediaiSD1}. 
Another hallmark of electron correlation is the Coulomb blockade phenomenon, 
which has been experimentally observed in organic molecular junctions~\cite{parkNat407,kubatkinNat425,pootNL6,aleshinPRB72,xuJACS127,songNat462} 
and graphene nanoribbons~\cite{hanPRL98,solsPRL99},
in the regime where the charging energy dominates over the molecule-electrodes coupling. 
In contrast, molecules in open-shell configurations are expected to exhibit significant electron–electron scattering within the partially filled SOMO.
%Indeed, 
In computational quantum chemistry, it is well-established that open-shell molecular configurations 
require careful treatment (see, e.g.,~\cite{krylov2017} for an overview) 
but the accuracy of quantum chemical methods comes at a prohibitively high numerical cost. 
Despite significant advances in simulation schemes 
for complex molecular and nano-electronic devices~\cite{jacobPRB82,jacobJPCM27,appeltNano10,droghettiPRB105,droghettiPRB106,valliNL18,valliPRB100,lupoNatCompSci1}, 
investigations of organic radicals within genuinely many-body frameworks 
(i.e., beyond static mean-field) remain limited. 
Moreover, they mostly focus either on Kondo physics~\cite{scottACSNano4,requistPNAS111,appeltNano10,mitchellNC8,jacobPRB104,mitraChem2025,li2408.01236}\textemdash relevant at low (cryogenic) temperatures, 
or on understanding inter-molecular exchange couplings~\cite{jacobPRB106,duNatComm14,zuoJPCL15,kraneNL20}, 
%In contrast, 
whereas the high-temperature phenomenology (i.e., above the Kondo scale) 
of stable and persistent organic radicals remains largely unexplored.

In the following, we focus on the electron transport 
in a two-terminal single-molecule junction configuration, 
in which an organic radical bridges metallic electron reservoirs. 
Specifically, we select a linear (pentadienyl) and a cyclic (benzyl) radical 
as archetypal representatives, 
since both molecules are $\pi$-radicals with a single unpaired electron, 
and natural candidates to manifest many-body effects. 

Our key result is to show that the Coulomb repulsion within the SOMO 
induces a \textit{many-body splitting} of the SOMO resonance. 
We suggest this scenario as a possible explanation for the relatively low conductance 
observed in transport experiments. 
In particular, we 
(i) identify the splitting as the hallmark of strong electron correlations, 
(ii) pinpoint the origin of the splitting, 
ascribed to a \textit{giant} electron scattering rate within the SOMO, and 
(iii) demonstrate that the phenomenology of the splitting 
is a \textit{non-perturbative} effect of the Coulomb repulsion, 
and cannot be obtained with less sophisticated techniques, 
such as perturbative approaches or static mean-field theories.  

We argue that the splitting of the SOMO is a general phenomenon that applies 
to a broad family of open-shell molecular systems, but at the same time, 
we highlight how a different spatial distribution of the SOMO 
bears dramatic consequences for the transport properties of the junction. 
Finally, we discuss its relevance in transport experiments 
and suggest a non-linear current-voltage characteristics 
as a possible experimental probe to confirm our predictions.

\section{Methods}\label{sec:methods}

We have developed a comprehensive numerical workflow that combines 
density functional theory (DFT) with quantum field theoretical methods, 
and is therefore able to address the chemical complexity of the radical molecule \emph{ab-initio} 
as well as electronic correlation effects beyond the single-particle picture. 
In the following, we discuss each building block and how they come together 
to provide a comprehensive and reliable framework to investigate 
the electronic and transport properties of strongly correlated molecule junctions.

\subsection{Local orbitals and low-energy models} \label{subec:los}

The art of combining \textit{ab-initio} and many-body approaches is to find a minimal orthogonal basis set,
which describes the correlated manifold. 
Maximally localized Wannier functions (WFs)~\cite{marzariRMP84} are routinely used in this context. 
However, due to their variational nature, WFs are not ideal in the context of quantum transport, 
as they require separate calculations for the scattering region and the electrodes. 
To date, there has been an ongoing effort to ease key challenges, including, 
for example, the automated generation of WFs~\cite{vitalenpjCM6} and interface matching~\cite{stieger2007.04268}. 
Variations of WFs, particularly suitable for molecules, have also been proposed~\cite{thygesenPRL94}. 

We follow an alternative approach relying on a linear combination of atomic orbitals (LCAO). 
At the heart of the proposed computational scheme lies a unitary transformation 
from the non-orthogonal atomic orbitals (AOs) used in the DFT calculation to a basis 
that reflects the local chemical environment of each atom in the molecule~\cite{borgesJCP144,gandusJCP153}. 
Following~\cite{gandusJCP153}, we refer to the rotated basis as \textit{local orbitals} (LOs).
So far, the LO approach has been employed in the context of DFT, 
for instance, to separate the contributions of the $\sigma$ and $\pi$ systems to the transmission function~\cite{borgesJCP144,borgesNL17,borgesJPCL7}, 
and to perform electron transport simulations very efficiently, 
without sacrificing numerical accuracy~\cite{gandusJCP153}. 
In what follows, we rely on the LOs approach in combination with quantum embedding 
to map the original Hamiltonian in the AO basis to an effective many-body problem 
in which the correlated active space, 
with hopping and Coulomb integrals evaluated in the LO basis, 
is embedded in a space representing the rest of the system. 
The effective problem can then be solved with appropriate numerical methods. 
This framework provides a comprehensive theoretical tool with predictive power, 
particularly suited for addressing the transport properties of molecular and nano-junctions.

The starting point is a DFT calculation in an AO basis set, within the LCAO scheme, leading to a  finite set of non-orthogonal Kohn-Sham orbitals $\{\ket{i}\}$ that span the Hilbert space $H$, with overlap matrix $\braket{i}{j}=(\mathbf{S})_{ij}\neq \delta_{ij}$ for $\ket{i}, \ket{j} \in H$. 
A set of LOs $\{\ket{m}\}$, with $\ket{m} \in M \subseteq H$ can be obtained for any atom $\alpha$ in subspace $M$ 
by a subdiagonalization of the corresponding Hamiltonian sub-block 
\begin{equation} \label{eq:HA}
    \mathbf{H}_{\alpha} \ket{m} = \epsilon_{m} \mathbf{S}_{\alpha} \ket{m}. 
\end{equation}
The LOs are therefore linear combinations of AOs and are, by construction, orthogonal on each atom. 
They can be classified according to symmetry, devoid of the redundancy characteristic of split-valence basis sets. 
This enables a more natural physical interpretation of the LOs as atomic orbitals~\cite{gandusJCP153}.
%The goal is to obtain an effective low-energy model which operates on a reduced part of the Hilbert space, but embeds the information of the environment. 
%To this end, we select a subset of LOs $\{\ket{a}\}=A \subset M$ 
%which are expected to describe the relevant physics close to the Fermi energy and separate it from the rest of the orbitals, or embedding, $\{\ket{e}\} \in E \equiv H \setminus A $. We then perform a L\"owdin orthogonalization~\cite{lowdin1950non} of the $\ket{a}$ states and redefine the $A$ subspace in terms of this new orthonormal basis set 
%with elements
%\begin{equation}
% \ket{a^\perp} = \sum_a(\mathbf{S}^{-1/2})_{aa^\perp} \ket{a}
%\end{equation}

To obtain an effective model \emph{ab-initio}, 
we formally separate the Hilbert space into an active space (A) and an environment (E). 
The active space consists of a subset of LOs $\{\ket{a}\}$, with $\ket{a} \in A \subseteq M$ 
which are expected to describe the %relevant 
physics close to the Fermi energy, 
and, at the same time, can be feasibly treated within quantum many-body techniques. 
The environment consists of all the remaining LOs and AOs $\{\ket{e}\}$, with $\ket{e} \in E \equiv H \setminus A$. 
Embedding the active space into the environment (rather than neglecting it) 
ensures that the effective model retains all information  
from the original single-particle Hamiltonian~\cite{gandusJCP153}. %as we shall see below. 
Finally, it is convenient to perform a L\"owdin orthogonalization~\cite{lowdin1950non} 
of the set of LOs $\{\ket{a}\}$, 
so that the active region $A$ is spanned by an orthonormal basis with elements 
\begin{equation}
 \ket{a^\perp} = \sum_a(\mathbf{S}^{-1/2})_{aa^\perp} \ket{a}.
\end{equation}
Since the overlap between LOs belonging to different atoms is typically small, 
i.e., $(\mathbf{S})_{ij} = \braket{m_i}{m_j}\ll 1$, 
the L\"owdin orthonormalization of the active space introduces only a weak deformation of the original LOs, 
thus preserving their (approximate) atomic-like symmetry. 
%\AV{[show it in SI in revision to submit to journal...]}

%\begin{equation}
% A \overset{\operatorname{def}}{=} \qty\big{\ket{a^\perp} \mid a^\perp=\sum_a(\mathbf{S}^{-1/2})_{aa^\perp} \ket{a}}
%\end{equation}

%The low-energy model is constructed for the $A$ subspace by integrating-out the orbitals of the $E$ subspace through a downfolding procedure~\cite{pavarini2004mott,solovyev2004lattice}. 
In practice, the (L\"owdin) LO low-energy model is constructed by embedding the active subspace into the environment through a downfolding procedure~\cite{pavarini2004mott,solovyev2004lattice}.
Taking into account the non-orthogonality between the $A$ and $E$ subspaces~\cite{jacobJPCM27}, 
the Green's function projected onto the $A$ subspace is given by
\begin{equation} \label{eq:GA}
    \mathbf{G}_A(z) = \mathbf{S}^{-1}_{A}\mathbf{S}_{AH} \mathbf{G}_H(z) \mathbf{S}_{HA}\mathbf{S}^{-1}_A,
\end{equation}
where $z=E+i\eta$ is a complex energy with 
%an infinitesimal imaginary shift $\eta$. 
an infinitesimal shift $\eta \rightarrow 0^{+}$. $\mathbf{G}_H$ denotes the Green's function of the full Hilbert space, and $\mathbf{S}_{AH}$ the overlap matrix between orbitals $\ket{a^\perp} \in A$ and orbitals $\ket{i} \in H$. 
Since the overlap $\mathbf{S}_A$ between the $\ket{a^\perp}$ states is the identity matrix by construction, it will be omitted in what follows for notational simplicity.
The effect of the environment on the $A$ subspace is described by the hybridization function 
\begin{equation} \label{eq:hybridA}
    \mathbf{\Delta}_A(z) = \mathbf{g}_A^{-1}(z) - \mathbf{G}_A(z)^{-1},
\end{equation}
where
\begin{equation} \label{eq:g0A}
    \mathbf{g}_A = \big[ z-\mathbf{H}_A \big]^{-1}
\end{equation}
is Green's function of the isolated $A$ subspace. Rewriting $\mathbf{G}_A$ in terms of $\mathbf{\Delta}_A$ and using the definition of $\mathbf{g}_A$ yields
\begin{equation} \label{eq:GA_eff}
    \mathbf{G}_A(z) = \big[ z - \mathbf{H}_A - \mathbf{\Delta}_A(z) \big]^{-1}.
\end{equation}
%which reveals that 
Then, $\mathbf{G}_A$ can be seen as the resolvent operator for an effective $A$ subspace 
renormalized by the environment through a dynamical hybridization. 
The Green's function describes the physics of the whole system, projected onto the active subspace. 
%If the states of the embedding and the active space are energetically well separated, 
%one can construct an effective Hamiltonian as
%\begin{equation} \label{eq:HA_eff}
%    \mathbf{H}^{\mathrm{eff}}_A = \mathbf{H}_A + \mathbf{\Delta}_A(z=0),
%\end{equation}
%where the energy dependence of the hybridization has been, to a good approximation, neglected. 

For a single-particle Hamiltonian, the partition above is arbitrary, and the procedure remains valid regardless of the subset of LOs included in the active space. 
In the context of $\pi$-conjugated organic molecules, 
the projection onto a single LO with p$_z$ symmetry per C atom 
(and, where relevant, additional species such as N or S) 
is usually sufficient to achieve a faithful representation of the frontier MOs, 
This choice is hence suitable to describe the physics close to the Fermi energy~\cite{gandusJCP153}, 
but more LOs can be included if required. 
The possibility of considering a restricted subset of LOs in the effective model is of pivotal importance 
in view of performing computationally-heavy many-body simulations.

\subsection{\textit{cRPA and ab-initio} Coulomb parameters}
In order to derive the electronic interaction parameters in the $A$ subspace beyond the %(semi-) 
semi-local density approximations, we employ the %so-called 
constrained Random Phase Approximation (cRPA)~\cite{jacobJPCM27,miyake2009ab,aryasetiawan2006calculations}. 
Within the cRPA, we select a region $R \supset A$ %surrounding the $A$ subspace 
where the formation of electron-hole pairs is expected to screen the Coulomb interaction between the $A$ electrons. 
Due to the strong local nature of the LOs, it is sufficient that $R$ comprises the $A$ subspace and a few atoms nearby~\cite{jacobJPCM27}. 
Defining $\mathbf{G}_R$ to be the Green's function projected onto the $R$ subspace, 
in analogy with Eq.~(\ref{eq:GA}), the screened Coulomb interaction is given by
\begin{equation} \label{eq:W_R} 
    \mathbf{W}_R(z) = \big[ \mathbf{I}-\mathbf{V}_R\mathbf{P}_R(z) \big] ^{-1}\mathbf{V}_R,
\end{equation}
where $\mathbf{V}_R$ is the bare Coulomb interaction 
\begin{equation}
    \mathbf{V}_{R; ij,kl} = \int \!\! dr \int \!\! dr' \psi^{\phantom{\dagger}}_i(r) \psi^{*}_j(r) 
                             \frac{e^2}{\abs{r-r'}} 
                             \psi^{*}_k(r') \psi^{\phantom{\dagger}}_l(r'),
\end{equation}
with $\psi_i(r)$ the wavefunction of the corresponding orbital in the $R$ region,
and $\mathbf{P}_R$ the polarizability at the RPA level 
\begin{equation} \label{eq:P_R}
    \mathbf{P}_{R;ij,kl}(z) = -2i \int \frac{dz'}{2\pi} \mathbf{G}_{R;ik}(z-z') \mathbf{G}_{R;lj}(z').
\end{equation}
The projection of $\mathbf{W}_R$ onto the $A$ subspace then yields 
the screened interaction $\mathbf{W}_A$. 
%Since we aim at obtaining the effective Coulomb parameters free of the self-interaction between $A$ electrons, we need to eliminate from $\mathbf{W}_A$ the screening channels arising from $A$-$A$ transitions included in $\mathbf{P}_R$, which are to be included in a more sophisticated treatment of the low-energy model.
%This is obtained with the following "unscreening" \AV{prescription}
Since we aim at performing many-body simulations of the effective model, we need to partially unscreen the Coulomb parameters by eliminating from $\mathbf{W}_A$ the screening channels arising from $A$-$A$ transitions included in $\mathbf{P}_R$. These contributions will instead be treated at a more sophisticated level of theory. 
This can be done according to the following prescription
\begin{equation} \label{eq:U_A}
    \mathbf{U}_A(z) = \mathbf{W}_A(z) \big[ \mathbf{I}+\mathbf{P}_A(z)\mathbf{W}_A(z) \big] ^{-1},
\end{equation}
using the polarization $\mathbf{P}_A$ of the $A$ electrons obtained from $\mathbf{G}_A$ similarly to Eq.~(\ref{eq:P_R}). 
The static components $\mathbf{U}_A(z=0)$ %can therefore be regarded as the effective 
are the (partially screened) Coulomb parameters in the active space.  
%In the following, we solve a low-energy model where the electron-electron interactions 
%are defined by retaining only this static component. 

\subsection{Solutions of the low-energy models}\label{sec:many-body}

The Green's function of Eq.~(\ref{eq:GA_eff}), 
and the interactions parameters of Eq.~(\ref{eq:U_A}), 
define a low-energy model in the LO basis that can be solved with many-body techniques. 
Here, we propose two somewhat complementary strategies based on exact diagonalization (ED) 
and dynamical mean-field theory (DMFT)~\cite{georgesRMP68} 
as implemented within its real-space generalization (R-DMFT) 
for inhomogeneous systems~\cite{snoekNJP10,valliPRL104,valliPRB92,jacobPRB82}. 

\subsubsection{Exact diagonalization}
%Since the ED solution technique requires a Hamiltonian formulation of the problem, 
%we obtain the effective model approximating the hybridization function with its static component, as defined in Eq.~(\ref{eq:HA_eff})
%and including the partially screened Coulomb interaction, thus yielding
%The starting point of ED is the many-body Hamiltonian in the basis of the effective model defined by Eq.~(\ref{eq:HA_eff}) including Coulomb interactions
The ED technique requires a Hamiltonian formulation of the effective model.  
%If the states of the active and embedding subspaces are energetically well-separated, 
%it is possible to neglect the dynamical character of the hybridization function 
A possibility is to neglect the dynamical character of the hybridization function 
and construct an effective Hamiltonian as
\begin{equation} \label{eq:HA_eff}
    \mathbf{H}^{\mathrm{eff}}_A = \mathbf{H}_A + \Re \mathbf{\Delta}_A(z=0),
\end{equation} 
where the hybridization term accounts for a renormalization 
of the hopping amplitudes and orbital energies within the active space due to the embedding.  
This is a reasonable approximation if the states of the active and embedding subspaces are energetically well-separated 
and/or the density of states of the embedding subspace is smooth around the Fermi energy. 
The model Hamiltonian then reads
\begin{equation}
\begin{split}
H & = \sum_{ij,\sigma} \big[ \mathbf{H}^{\mathrm{eff}}_A - \mathbf{H}^{\mathrm{dc}}_A \big]_{ij} c^{\dagger}_{i\sigma} c^{\phantom{\dagger}}_{j\sigma} \\ 
  & + \frac{1}{2} \sum_{ijkl,\sigma\sigma'} \mathbf{U}_{A;ij,kl} c^{\dagger}_{j\sigma} c^{\dagger}_{k\sigma'} c^{\phantom{\dagger}}_{l\sigma'} c^{\phantom{\dagger}}_{i\sigma},
 \end{split}
\end{equation}
where $c^{(\dagger)}_{i\sigma}$ denotes the annihilation (creation) operator 
of an electron at LO $i$ with spin $\sigma$, 
$\mathbf{U}_A$ is the matrix of the cRPA screened Coulomb interaction parameters, 
and the double-counting correction $\mathbf{H}^{\mathrm{dc}}_A$ accounts for the interaction 
already included at the mean-field level within DFT (see Sec.~\ref{sec:dc} for the details).  
%The diagonalization of this Hamiltonian yields the many-body eigenstates and eigenvalues which can be used to construct the single-particle Green's function $\mathbf{G}^{\mathrm{ED}}_A$ \footnote{The term ``single-particle'' must not be confused with the underlying level of theory, since $\mathbf{G}^{ED}_A$ is truly a many-body object describing the propagation of an electron in the full many-body Hamiltonian.} making use of the Lehmann spectral representation \cite{senechal2008introduction,dolfen2007massively}. The correlated self-energy can then be obtained from the Dyson equation
The diagonalization of this Hamiltonian yields its many-body spectrum (i.e., eigenorbitals and eigenvalues) which can be used to construct the Green's function $\mathbf{G}^{\mathrm{ED}}_A$ through its Lehmann representation~\cite{bruus2004many}.
The many-body self-energy is obtained from the Dyson equation
\begin{equation} \label{eq:SigmaED}
    \mathbf{\Sigma}^{\mathrm{ED}}_A(z) = z - \mathbf{H}^{\mathrm{eff}}_A-\big[\mathbf{G}^{\mathrm{ED}}_A(z)\big]^{-1},
\end{equation}
and it describes local $\mathbf{\Sigma}_{A;ii}$ 
and non-local $\mathbf{\Sigma}_{A;i \neq j}$ electronic correlations 
in the LO basis. 
While ED is limited to $\mathcal{O}(20)$ orbitals, 
an obvious advantage over approaches such as quantum Monte Carlo~\cite{gullRMP83}, 
is that ED provides direct access to the retarded self-energy and Green's function 
(and therefore also to the transmission function) 
without the need to perform a numerical analytic continuation, 
which is known to be an ill-defined problem~\cite{jarrellPR269}. 
Note that within ED, we obtain a many-body self-energy which is, by construction, spin-independent, i.e., 
$\Sigma^{\sigma}_{ij}=\Sigma^{\bar{\sigma}}_{ij}$ 
since $\mathbf{H}^{\mathrm{eff}}_A$ follows from a restricted DFT calculation.

\subsubsection{Real-space DMFT}
The idea behind R-DMFT consists of mapping the many-body problem 
onto a set of auxiliary Anderson impurity models (AIMs) ---one for each atom $\alpha$--- 
described by the projected Green's function~\cite{snoekNJP10,valliPRL104,valliPRB92} 
\begin{equation} \label{eq:auxAIM}
    \mathbf{g}^{\sigma}_{\alpha}(z) = \left[ \mathbf{G}^{\sigma}_A(z) \right]_\alpha.
\end{equation}
The solution of AIM $\alpha$ yields 
a local many-body self-energy $\mathbf{\Sigma}^{\sigma}_{\alpha}(z)$. 
Therefore, the self-energy of the $A$ subspace is block diagonal in the atomic subspaces   
\begin{equation} \label{eq:SigmaDMFT_A}
    \mathbf{\Sigma}^{\sigma}_{A}(z) = \mathrm{diag}(\qty\big{\mathbf{\Sigma}^{\sigma}_{\alpha}(z) \mid \alpha \in A}).
\end{equation}
The auxiliary AIMs are coupled by the Dyson equation
\begin{equation} \label{eq:GDMFT_A}
    \mathbf{G}^{\sigma}_A(z) = \big[ z + \mu - (\mathbf{H}_A - \mathbf{H}^{\mathrm{dc}}_A) 
                             - \mathbf{\Delta}_A(z) - \mathbf{\Sigma}^{\sigma}_A(z) \big]^{-1},
\end{equation}
where the Green's function $\mathbf{G}^{\sigma}_A$ includes the many-body self-energy and the double-counting correction (see Sec.~\ref{sec:dc} for the details). 
The chemical potential $\mu$ can either be determined self-consistently 
to fix the total occupation of the active space (e.g., to preserve the DFT occupation of the $A$ subspace) 
or set to a predefined value. 
%We choose to set $\mu=0$ so that the Fermi energy is determined by the electron reservoirs in the junction 
%(which is the reference energy of the AO Hamiltonian). 
%Consequently, any shift in the orbital energies in the many-body calculation 
%results in a charge transfer between the active space and its embedding environment. 
Finally, Eqs.~(\ref{eq:auxAIM}-\ref{eq:GDMFT_A}) are iterated self-consistently 
starting from an initial guess 
(typically the uncorrelated solution with $\mathbf{\Sigma}^{\sigma}_{A}=0$) 
until convergence. 

In more detail, in AIM $\alpha$ the impurity electrons interact 
through a screened local Coulomb repulsion projected onto atom $\alpha$, 
i.e., $\mathbf{U}_{\alpha} = \mathbf{U}_{A;ij,kl} \mid \{i,j,k,l\} \in \alpha$~\footnote{This approximation neglects non-local interaction terms, which could otherwise be taken into account either at the mean-field level or within alternative implementations, such as extended DMFT~\cite{sunPRB66}.}.
Moreover, the impurity is embedded in a self-consistent \emph{bath} of non-interacting electrons, which describes the rest of the electronic system, encoded in the hybridization function 
\begin{equation} \label{eq:DeltaDMFT_A}
    \mathbf{\Delta}^{\sigma}_{\alpha}(z) = z + \mu - (\mathbf{H}_{\alpha} - \mathbf{H}^{\mathrm{dc}}_{\alpha}) - \big[\mathbf{g}^{\sigma}_{\alpha}(z)\big]^{-1} 
                                         - \mathbf{\Sigma}^{\sigma}_{\alpha}(z). 
\end{equation}
Also within R-DMFT, it is convenient to use an ED impurity solver for the AIMs 
to have direct access to retarded functions. 
This requires a \emph{discretization} of the hybridization function with a finite number of bath orbitals, described by orbital energies $\epsilon^{\sigma}_{m}$ 
and hopping parameters to the impurity orbital $t^{\sigma}_{mi}$. 
The hybridization parameters together with the local Coulomb blocks $\mathbf{U}_\alpha$, 
define the AIM Hamiltonian 
\begin{equation}
\begin{split}
    H_{\mathrm{AIM}} & = \sum_{ij,\sigma} \big[ \mathbf{H}_\alpha-\mathbf{H}^{\mathrm{dc} }_\alpha \big]_{ij} c^{\dagger}_{i\sigma} c^{\phantom{\dagger}}_{j\sigma} - \mu \sum_{i\sigma} c^{\dagger}_{i\sigma} c^{\phantom{\dagger}}_{i\sigma} \\
                     & + \sum_{m,\sigma} \epsilon^{\sigma}_{m} a^{\dagger}_{m\sigma} a^{\phantom{\dagger}}_{m\sigma} 
                       + \sum_{mi,\sigma} t^{\sigma}_{mi} (a^{\dagger}_{m\sigma} c^{\phantom{\dagger}}_{i\sigma} + c^{\dagger}_{i\sigma} a^{\phantom{\dagger}}_{m\sigma}) \\
                     & + \frac{1}{2} \sum_{ijkl,\sigma\sigma'} \mathbf{U}_{\alpha;ij,kl} 
                         c^{\dagger}_{j\sigma} c^{\dagger}_{k\sigma'} c^{\phantom{\dagger}}_{l\sigma'} c^{\phantom{\dagger}}_{i\sigma},
\end{split}
\end{equation}
where $c^{(\dagger)}_{i\sigma}$ and $a^{(\dagger)}_{m\sigma}$ denote the annihilation (creation) operator of an electron at LO $i \in \alpha$ with spin $\sigma$, or at bath orbital $m$ with spin $\sigma$, respectively. 
Once the many-body spectrum of the AIM is known, the local self-energy is evaluated in terms of the local Green's function $\mathbf{G}^{\sigma}_{\alpha}$ as 
\begin{equation}
    \mathbf{\Sigma}^{\sigma}_{\alpha}(z) = \big[\mathbf{g}^{\sigma}_{\alpha}(z)\big]^{-1}-\big[\mathbf{G}^{\sigma}_{\alpha}(z)\big]^{-1}, 
\end{equation}
and $\mathbf{\Sigma}^{\sigma}_A$ according to Eq.~(\ref{eq:SigmaDMFT_A}). 
At convergence, we define the R-DMFT self-energy as 
\begin{equation} \label{eq:SigmaDMFT}
    \mathbf{\Sigma}^{\sigma,\mathrm{R-DMFT}}_A(z) = \mathbf{\Sigma}^{\sigma}_A(z) - \mathbf{H}^{\mathrm{dc}}_A - \mu,
\end{equation}
so that it includes all shifts associated with the density matrix.

In terms of approximations, R-DMFT accounts for local electronic correlations ($\mathbf{\Sigma}_{A;ii}$), 
while neglecting non-local correlations (i.e., $\mathbf{\Sigma}_{A;i \neq j}=0$). 
However, some degree of non-locality is retained since $\mathbf{\Sigma}_{A;ii}\neq\mathbf{\Sigma}_{A;jj}$, 
as the AIMs are coupled self-consistently through the Dyson equation. R-DMFT is therefore suitable to describe electronic correlations in intrinsically 
inhomogeneous systems~\cite{valliPRB86,dasPRL107,valliPRB92,kropfPRB100,baumannPRA101,jacobPRB82}.
Moreover, the computational complexity of R-DMFT is 
exponential in the number of LOs in each atom $\alpha$ but only linear in the number of atoms. 
Since the AIMs are defined by Eq.~(\ref{eq:auxAIM}) 
are \textit{independent}, this also allows 
(i) exploiting any real-space symmetry of the many-body Hamiltonian,  
and solve only the \textit{locally inequivalent} AIMs, and
(ii) parallelizing the solution of said impurity problems. 
Therefore, R-DMFT is considerably lighter than direct ED of the original many-body problem in terms of computational complexity,  
and can treat systems with hundreds of atoms in the active space~\cite{snoekNJP10,valliPRB92,valliPRB86}.
% In terms of approximations, R-DMFT takes into account local electronic correlations ($\Sigma_{ii}$).
% While non-local correlations are neglected (i.e., $\Sigma_{ij}=0$),
% some degree of non-locality is retained as $\Sigma_{ii}\neq\Sigma_{jj}$ 
% and the AIMs are coupled through the self-consistent Dyson equation. 
% Therefore, R-DMFT is suitable to treat intrinsically 
% inhomogeneous systems~\cite{valliPRB86,dasPRL107,valliPRB92,kropfPRB100,baumannPRA101,jacobPRB82}. 
% Moreover, the computational cost of R-DMFT scales as ${\cal O}(N_{\alpha} \times 4^{N_{m}})$, 
% with $N_{\alpha}$ and $N_m$ being the number of atoms in the active space and the number of LOs per atom, respectively, 
% and it is significantly reduced with respect to the direct ED of the original many-body problem, 
% which is ${\cal O}(4^{N_{\alpha}\times N_m})$. 
% Thus, R-DMFT can easily treat systems with hundreds of atoms in the active space, inaccessible to ED~\cite{snoekNJP10,valliPRB92,valliPRB86}. 
Finally, beyond the restricted solution $\mathbf{\Sigma}_A^{\sigma}=\mathbf{\Sigma}_A^{\bar\sigma}$, 
R-DMFT also permits breaking spin degeneracy 
and thus describing magnetic solutions with short-range order~\cite{snoekNJP10,valliPRB94,valliNL18,valliPRB100,amaricciPRB98}.

\subsection{Double-counting correction} \label{sec:dc}
The double-counting (DC) correction $\mathbf{H}^{\mathrm{dc}}_A$ aims at eliminating 
the electron-electron correlations in the $A$ subspace 
accounted for at mean-field level in the DFT calculation, 
which will be replaced by those obtained at a more sophisticated level of theory 
in the many-body simulations. %treatment of the low-energy model. 
Unfortunately, %we don't know exactly which correlation effects are accounted for in DFT, 
since DFT (in contrast to, e.g., GW) does not have a formulation in terms of Feynman diagrams, 
the analytical expression of such a correction within DFT is unknown. 
Several approximations~\cite{anisimov1991band,czyzyk1994local,karolak2010double,jacobPRB82} have been developed in the context of DFT+U~\cite{anisimov1997first,petukhov2003correlated} 
or DFT+DMFT~\cite{kotliar2006electronic,held2007electronic}.  
%For a single-orbital AIM (as in the case of the simulations in this work) a well-established DC correction can be reasonably approximated within the fully localized limit (FFL)~\cite{liechtenstein1995density,anisimov1993density,czyzyk1994local,solovyev1994corrected} 
Here, we choose the fully localized limit (FLL)~\cite{liechtenstein1995density,anisimov1993density,czyzyk1994local,solovyev1994corrected} 
generalized to the case of an anisotropic Coulomb repulsion~\cite{jacobJPCM27}. 
%which for a single-orbital AIM (as in the case of the simulations in this work) reads 
%
%\begin{equation} \label{eq:FLLdc}
%    \mathbf{H}^{\mathrm{dc}}_{A;ii} = \sum_{j=1}^{N_{\mathrm{orb}}} \mathbf{U}_{A;ii,jj} \left(n^{\mathrm{DFT}}_j - \frac{1}{2 N_{\mathrm{orb}}} \right)
%                                      + \frac{1}{2} \left( \sum_j U_{A;ijji} \, n^{\mathrm{DFT}}_j - 1 \right),
%\end{equation}
%with intra-atomic orbital indices $i, j \in \alpha$, 
For a single correlated orbital per atom in the correlated subspace (as in the case of the simulations in this work) this reduces to 
\begin{equation} \label{eq:FLLdc-so}
    \mathbf{H}^{\mathrm{dc}}_{A;ii} = \mathbf{U}_{A;ii,ii} \left(n^{\mathrm{DFT}}_i - \frac{1}{2} \right),
\end{equation}
where $n^{\mathrm{DFT}}_i$ is the DFT occupation of orbital $i$. 
%\AV{[this is not completely general, because in the multi-orbital case there is an additional term $\frac{J}{2}(1-n_{ii}^{\mathrm{DFT}})$ and in this notation $J = (\mathbf{U}_A)_{ij,ji}$]}
Hence, we use this form of DC for the R-DMFT calculations. 
However, there's no established method for the general case of multi-site and multi-orbital Coulomb interaction 
as encountered in ED. 
A possible choice is to require the correlated self-energy to fulfill the condition~\cite{karolakJESRP181}
\begin{equation}
    \Re \mathrm{Tr}\left[\mathbf{\Sigma}_{A}(|z|\to\infty)\right] = 0,
\end{equation}
which we implement here within a self-consistent procedure 
optimizing a set of local parameters that play the role 
of effective orbital occupations in Eq.~(\ref{eq:FLLdc-so}). 
This approach ensures that the electronic properties at high energies, which are well described by a one-particle approach, 
are restored (on average) to the DFT level. 
% which ensures that the local crystal fields are restored at the DFT level.

\subsection{Correlated quantum transport} \label{sec:corrQT}
To describe the electronic transport properties, we use the non-equilibrium Green's function (NEGF) approach \cite{datta2005quantum,ryndyk2016theory}. 
In NEGF, we identify a device region that extends beyond the molecular bridge, 
and downfold the leads' electrons by virtue of an efficient recursive algorithm~\cite{gandusSSE199}. 
The corresponding Green's function reads  
\begin{equation}
    \mathbf{G}_D(z) = \big[  z\mathbf{S}_D - \mathbf{H}_D 
                            - \mathbf{\Sigma}_L(z) - \mathbf{\Sigma}_R(z) -\mathbf{\Sigma}_D(z) \big]^{-1},
\end{equation}
where $\mathbf{\Sigma}_{L(R)}$ is the self-energy describing the electrons in the left (right) electrodes, and
\begin{equation}
  \mathbf{\Sigma}_D(z) = \mathbf{S}_{DA} \mathbf{S}^{-1}_A \mathbf{\Sigma}_A(z) \mathbf{S}^{-1}_A \mathbf{S}_{AD}
\end{equation} 
projects the many-body self-energy of the active space $\mathbf{\Sigma}_A$ (i.e., obtained within either ED or R-DMFT) onto the device region.
Following the generalization of the Landauer formula proposed by Meir and Wingreen~\cite{meirPRL68}, the conductance is given by 
\begin{equation} \label{eq:conductance}
    G = G_0 T(E_F),
\end{equation}
where $G_0=e^2/h$ is the conductance quantum, and the transmission function is computed as
\begin{equation} \label{eq:transmission}
    T(E) = \Trace [\mathbf{\Gamma}_L(z) \mathbf{G}^{\dagger}_D(z) \mathbf{\Gamma}_R(z) \mathbf{G}_D(z)],
\end{equation}
with $\mathbf{\Gamma}_{L(R)}$ the anti-hermitian part of $\mathbf{\Sigma}_{L(R)}$
\begin{equation} \label{eq:broadening}
    \mathbf{\Gamma}_{L(R)}(z) = i \big[\mathbf{\Sigma}_{L(R)}(z) - \mathbf{\Sigma}^\dagger_{L(R)}(z) \big].
\end{equation}
Note that Eqs.~(\ref{eq:conductance})$-$(\ref{eq:broadening}) neglect the incoherent contributions 
(i.e., due to inelastic scattering) to the transmission 
that arise from the many-body self-energy~\cite{meirPRL68,ferrettiPRB72,ngPRL76,sergueevPRB65,nessPRB82,droghettiPRB105,droghettiPRB106}. 
In the non-resonant transport regime, and where electron-electron scattering is low, a Landauer transmission 
with renormalized Green's functions is expected to provide 
a good approximation of the transport properties at low bias voltage, 
even in the presence of strong correlations within the $A$ subspace~\cite{meirPRL68,jacobJPCM27}. 

% In general, however, the current has to be recast in terms of an effective transmission kernel, 
% where the Green’s function is dressed by a many-body self-energy 
% and the molecule-lead coupling is renormalized by vertex corrections~\cite{oguriJPSJ70,ferrettiPRB72,nessPRB82,droghettiPRB95}. 
% In the resonant transport regime, and therefore also for radicals, one can expect such contributions to become important. 
% Following~\cite{ferrettiPRB72}, the expression for the transmission function reads 
% \begin{equation} \label{eq:transmission}
%     T(E) = \Trace [\mathbf{\Gamma}_L(z) \mathbf{G}^{\dagger}_D(z) \mathbf{\Gamma}_R(z) \mathbf{\Lambda}(z) \mathbf{G}_D(z)],
% \end{equation}
% where 
% \begin{equation}
%     \mathbf{\Lambda}(z) = 1 + \big[ \mathbf{\Gamma}_L(z) + \mathbf{\Gamma}_R(z) \big]^{-1} \mathbf{\Gamma}_D(z)
% \end{equation}
% with 
% \begin{equation}
%     \mathbf{\Gamma}_D(z) = i \big[\mathbf{\Sigma}_{D}(z) - \mathbf{\Sigma}^\dagger_{D}(z) \big].
% \end{equation}

% Hence, in this picture, the vertex corrections correspond to an asymmetric renormalization of the molecule-lead coupling, 
% introducing an effective electrode that describes many-body scattering processes~\cite{ferrettiPRB72}. 

However, it can be shown that the transmission can be recast as
\begin{equation} \label{eq:transmission}
     T(E) = \Trace [\mathbf{\Gamma}_L(z) \mathbf{G}^{\dagger}_D(z) \mathbf{\Lambda}_R(z) \mathbf{G}_D(z)],
\end{equation}
so that correlation effects are included  
in a effective (renormalized) coupling to one of two electrodes 
\begin{equation}
    \mathbf{\Lambda}_R(z) = \mathbf{\Gamma}_R(z) + \mathbf{\Gamma}_c(z). 
\end{equation}
Hence, the transmission function it is naturally 
separated in two contributions: 
(i) a \textit{bubble} term that reduces to the Landauer transmission 
as $\mathbf{\Lambda}_R(z) \to \mathbf{\Gamma}_R(z)$, and 
(ii) a \textit{vertex correction} arising from $\mathbf{\Gamma}_c(z)$,
that takes into account inelastic electron scattering. 
The calculation of the vertex correction requires 
either the knowledge of a vertex function~\cite{oguriJPSJ70} 
or of the non-equilibrium distribution of the electrons 
in the presence of interactions~\cite{meirPRL68,ferrettiPRB72,ngPRL76,sergueevPRB65}.   
While this goes beyond the scope of the present work, we note that in the literature, 
different approximate recipes have been proposed to tackle this problem.~\cite{ngPRL76,oguriJPSJ70,nessPRB82,dosSantosNanotech36}.

\begin{figure}[bhp]
\includegraphics[width=1.0\linewidth, angle=0]{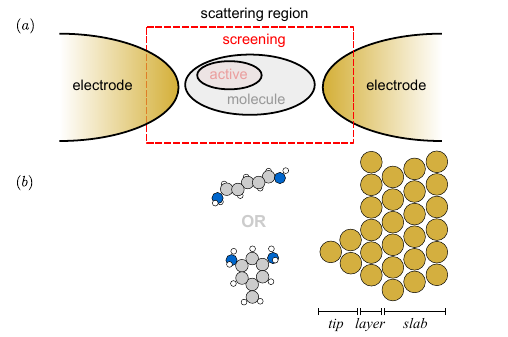}
\caption{(a) Schematics of the scattering region of the single-molecule junction, 
consisting of the molecular bridge and the Au electrodes. 
The screening space (R) and active space (A) 
intended as a subspace of the orbitals of the molecule, 
are highlighted. 
(b) Structure of the pentadienyl and benzyl radicals, 
substituted with the amino anchoring groups, and of the Au(111) electrodes. }
\label{fig:junctions_schematic}  
\end{figure}

\section{Computational Details}\label{sec:computational}
The structures were set up with the atomic simulation environment (ASE) software package~\cite{larsenJPCM29} 
and the DFT calculations were performed with the GPAW package~\cite{mortensenPRB71,larsenPRB80,enkovaaraJPCM22}. 
Geometry optimizations were performed by relaxing atomic positions until the forces on each atom were below $0.001$~Hartree/Bohr$^{-1}$ ($\approx 0.05$~eV/\AA). 
To converge the electron density, we used an LCAO double-$\zeta$ basis set, with a grid spacing of $0.2$~\AA,  
and the Perdew–Burke–Ernzerhof exchange-correlation functional~\cite{perdew_burke_ernzerhof1996}.
%For the electron transport calculations, we followed the method described in~\cite{gandusSSE199}. 
The leads were modeled by a three-layer-thick Au(111) slab 
sampled with a $3\times1\times1$ \textit{k}-point grid along the transport direction. 
The scattering region also includes one Au slab and an additional Au layer terminated by a four-atom Au tip, 
to which the molecule anchoring groups are attached.

For all structures considered below, the $A$ subspace describing the effective model consists of 
the p$_z$ LOs of the C and N atoms of the molecular bridge~\cite{gandusSSE199}, 
while the $R$ subspace for the cRPA calculation of the screened interaction  
includes the molecule and also extends to the Au atoms of the tip (see Fig.~\ref{fig:junctions_schematic}). 

The R-DMFT simulations were performed with and ED impurity solver~\cite{amaricciCPC273,crippaSPPCb58}
converging the solution 
with up to $n_b = 7$ bath sites discretization for each AIM. 
The ED simulations we performed using \texttt{edpyt}~\cite{edpyt}. 
In both cases we choose to set $\mu=0$ so that the Fermi energy is determined by the electron reservoirs in the junction (which is the reference energy of the AO Hamiltonian) 
and recalculate the occupation of the active space self-consistently.  
The quantum transport NEGF simulations without and with a many-body self-energy 
were performed using \texttt{qtpyt}~\cite{qtpyt}.

\section{Insights from ab-initio Simulations}
To understand the many-body effects arising in an open-shell configuration, 
it is useful to recall relevant chemical properties of the parent radicals  and show how those are reflected in their substituted counterparts 
in the junction setup from \emph{ab-initio} simulations. 
In particular, we inspect the spatial distribution of the SOMO 
and the cRPA Coulomb parameters projected onto the LOs active space. 

\begin{figure}[bp]
\includegraphics[width=1.0\linewidth, angle=0]{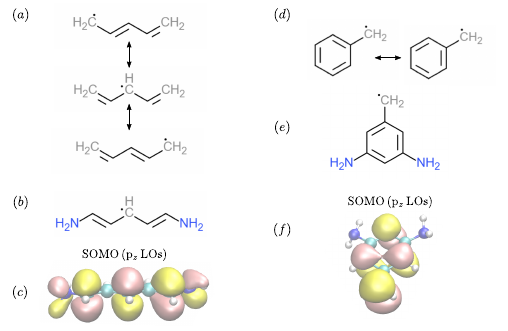}
\caption{(a,d) Resonant structures of the pentadienyl and benzyl radicals in the gas phase.
(b,e) Structure of the radical molecules substituted with amino anchoring groups. 
(c,f) SOMO isosurfaces projected on the p$_z$ LOs of the substitued molecules. 
Isovalues: $\pm 0.03$ au. See text for a discussion. 
%In pentadienyl, the unpaired electron is hosted by one of the \emph{odd} C of the polyene chain, 
%which also display the largest contributions in the isosurface, while the \emph{even} C correspond to nodes.  
%In benzyl, the unpaired electron is delocalized between the benzylic C and the benzene ring, along which the isosurface displays nodes on every other C.  
%A striking difference is that the SOMO in the pentadienyl structure has a sizable weight on the anchoring groups, while in the benzyl structure it displays nodes, thus suggesting a strong and a weak coupling to the electrodes, respectively. 
}
\label{fig:radicals_SOMO}  
\end{figure}

\subsection{Structure of the SOMO} \label{sec:SOMO}
The pentadienyl radical, with the chemical formula [CH$_2$(CH)$_3$CH$_2$]$^{z}$ (with total charge $z=0$), 
is a linear molecule and the shortest polyene radical after allyl. 
It has three resonant structures, in each of which the unpaired electron is hosted on one of the \emph{odd} C atoms. 
The delocalization of the unpaired electron along the molecular backbone contributes 
to the thermodynamic stability of the molecule~\cite{clark1991studies,chalyavi2011spectroscopy}. 
In the junction setup, the hydrogen atoms at each end of the chain are substituted 
by amino anchoring groups. 
By diagonalization of the AOs Hamiltonian, restricting to the subspace of the molecule, 
we obtain the corresponding fragment MOs, 
and we find an eigenvalue just above the Fermi energy, corresponding to the SOMO. 
The SOMO reflects the resonant structures, with its largest projection on the odd carbon atoms, nodes on the even ones, and a significant weight on the anchoring groups -- indicating strong coupling to the electrodes.
The pentadienyl resonant structures, the substituted molecule, 
and the projection of the SOMO onto the p$_z$ LOs of the active space 
are shown in Figs.~\ref{fig:radicals_SOMO}(a,b,c), respectively.

The benzyl radical, [C$_6$H$_5$CH$_2$]$^{z}$ (with total charge $z=0$), is derived from benzene (C$_6$H$_6$) by substituting a hydrogen atom with a methylene CH$_2$ group. The radical stabilized by resonance and the unpaired electron is delocalized between the benzylic carbon atom and the benzene ring.  
In the junction setup, the amino groups are substituted 
at the 1,3-positions of the aromatic ring 
(usually referred to as the \emph{meta} configuration in benzene) 
while the methylene group is substituted at the benzene 5-position, 
i.e., along the longer branch of the substituted ring. 
In this case, the SOMO displays its largest projection on the $p_z$ LO of the benzylic carbon atom, with alternating nodes on the ring carbons, but negligible weight on the anchoring groups, thus suggesting a weak coupling to the electrodes. 
The benzyl resonant structures, the substituted molecule, 
and the projection of the SOMO onto the p$_z$ LOs of the active space 
are shown in Figs.~\ref{fig:radicals_SOMO}(d,e,f), respectively.

%This observation is the key to understanding the drastically different transport properties of the two radical junctions.  
%The comparison highlights the key difference: while the pentadienyl SOMO couples strongly to the electrodes via the anchoring groups, %the benzyl SOMO, displays nodes at the anchoring groups.  
The comparison highlights a key difference: the pentadienyl SOMO has a sizable projection 
on the N-p$_z$ LOs of the amino groups, whereas in benzyl it displays nodes. 
This distinction anticipates the drastically different transport properties of the two radical junctions.

\begin{figure}[bp]
\includegraphics[width=\linewidth, angle=0]{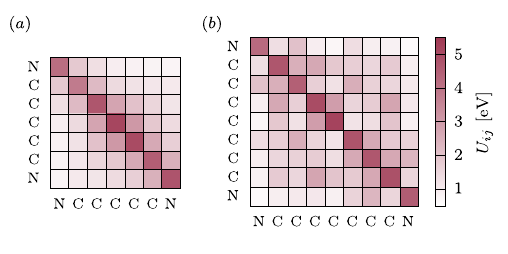}
\caption{Partially screened Coulomb parameters $U_{ij} = \mathbf{U}_{A;ij}$ in the LO basis 
for (a) pentadienyl and the (b) benzyl radicals. }
\label{fig:U_matrix}  
\end{figure}

\subsection{Coulomb parameters in the LO basis} \label{sec:UijLOs} 
The partially screened cRPA Coulomb matrix projected onto the LO basis of the active space 
is shown in Figs.~\ref{fig:U_matrix}(a,b) for the pentadienyl and the benzyl radicals, respectively. 
We adopt the simplified notation $U_{ij} = \mathbf{U}_{A;ij}$. 
In both cases, the intra-orbital Coulomb interaction parameters 
are in the range of $U_{ii} \approx 4$--$5$~eV 
and are slightly stronger for the atoms farther away from the metallic Au electrons, 
due to a weaker screening. 
Similar values of the Coulomb repulsion are found for the amino anchoring groups. 
However, as we shall discuss later in more detail, 
the C-p$_z$ LOs are generally close to half-filling, where correlations are the strongest, 
whereas the N-p$_z$ LOs are closer to being fully occupied, thus resulting in weaker correlations.

\section{Electron transport}\label{sec:transport}
In this section, we present the electron transport properties 
of single-molecule junctions with pentadienyl and benzyl derivative bridges.  
We compare the predictions of DFT+NEGF 
with approaches that take into account many-body effects at a level of approximation beyond DFT, 
and show that the two radicals exhibit very different electronic and transport properties. 
In particular, we show that different treatments of the spin degree of freedom result in \textit{qualitatively} different mechanisms for the splitting of the SOMO resonance.  
Which scenario applies in reality should be determined through experiments.

\subsection{Pentadienyl}\label{sec:pentadienyl}

\begin{figure}[tp!]
\includegraphics[width=1.0\linewidth, angle=0]{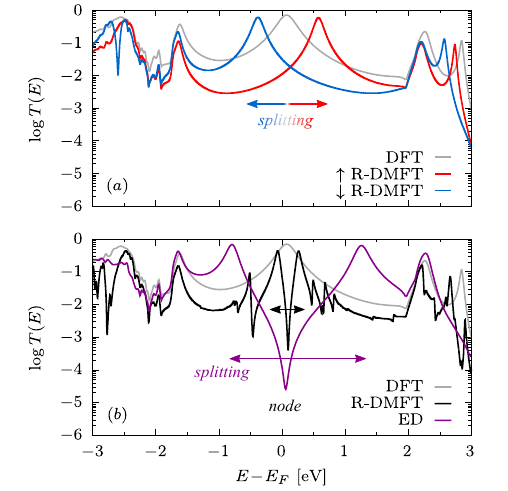}
\caption{Electron transmission function through the pentadienyl radical junction. 
DFT predicts a SOMO resonance close to $E_F$. Taking into account the Coulomb repulsion beyond restricted DFT yields: 
(a) a splitting of the resonance into $\downarrow$-SOMO and $\uparrow$-SUMO due to spin-symmetry breaking; 
(b) a splitting of the resonance due to many-body effects (without spin symmetry breaking) 
revealing a transmission node close to the Fermi energy.}
\label{fig:Te_pentadienyl}  
\end{figure}

Within DFT, the transmission function displays a resonance close to the Fermi energy (denoted by $E_F$) 
corresponding to ballistic transport through the SOMO. 
The resonance is found at $\epsilon_{\mathrm{SOMO}}= 70$~meV 
and has a width $\Gamma_{\mathrm{SOMO}}\approx 300$~meV, 
reflecting a significant 
hybridization of the SOMO with the states of the electrodes.
The slight misalignment between the SOMO resonance and $E_F$, 
yields a conductance $G=5.7 \times 10^{-1} \ G_0$ in each spin channel, see Fig.~\ref{fig:Te_pentadienyl}(a).
This scenario changes as the SOMO resonance is split due to the Coulomb repulsion. 
However, as we shall discuss below, depending on the splitting mechanism, we observe fundamentally different transport properties.

\textit{Symmetry-breaking.} In the spin-unrestricted R-DMFT simulations, 
the spin-doublet degeneracy is lifted explicitly by applying a symmetry-breaking field, 
which is then removed during the self-consistent loop. 
The SOMO position displays a spin-dependent shift, with   
an occupied state ($\downarrow$-SOMO) in the majority-spin channel and 
an unoccupied state ($\uparrow$-SUMO) in the minority-spin channel. 
Thus, the system realizes a magnetic insulator with a spin gap $\Delta_s \approx 1.0$~eV 
and a magnetic moment $\langle S_z \rangle \simeq 1/2$ due to the single unpaired electron. 
A corresponding resonance is found in each of the spin-resolved transmission functions, 
%as well as the sum over both spin channels (black dashed line) 
as shown in Fig.~\ref{fig:Te_pentadienyl}(a).
The spin-dependent shift is approximately symmetric around the Fermi level, 
yielding a similar conductance in the two spin channels 
$G^{\downarrow}=3.1 \times 10^{-2} \, G_0$ and $G^{\uparrow}=1.1 \times 10^{-2} \, G_0$ 
and low spin-filtering efficiency.  %$\eta \simeq 0.13$. 
Probing the transmission $T=T^{\uparrow}+T^{\downarrow}$ without spin resolution 
would result in a pair of resonances, associated with a SOMO-SUMO splitting 
and a conductance $G=G^{\uparrow}+G^{\downarrow}=4.2 \times 10^{-2} \, G_0$.
Importantly, deep in the symmetry-broken phase, 
the many-body self-energy displays a weak energy dependence 
(see also Sec.~\ref{sec:mechanism}) and similar results would be obtained 
by the static mean-field Hubbard~\cite{valliPRB94} or DFT+U approaches. 
Breaking the spin symmetry is a common approach often used in the literature, and the spin-dependent shift of a transmission resonance~\cite{herrmannJACS132,herrmann2011designing,smeuJPCC114} 
or antiresonance~\cite{valliNL18,valliPRB100}, 
has been suggested as a suitable mechanism for the realization of organic spin filters. 
%with the spin-filtering efficiency defined as
%\begin{equation}
% \eta = \bigg| \frac{G^{\uparrow}-G^{\downarrow}}{G^{\uparrow}+G^{\downarrow}} \bigg|.
%\end{equation}

\begin{figure}[tp!]
\includegraphics[width=1.0\linewidth, angle=0]{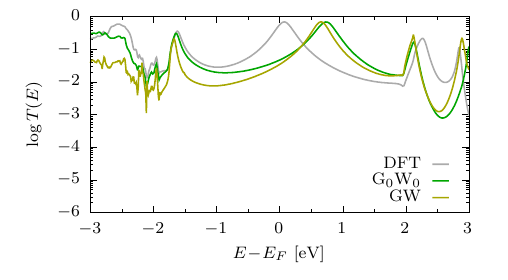}
\caption{Electron transmission function through the pentadienyl radical junction. 
Both the $G_0W_0$ and the self-consistent $GW$ approximations fail to predict the splitting of the SOMO feature, as described within ED and R-DMFT, cfr. Fig.~\ref{fig:Te_pentadienyl}. }
\label{fig:Te_GW}  
\end{figure}

\textit{Many-body splitting.} % The splitting of the SOMO can also be obtained 
The scenario is completely different within either spin-restricted R-DMFT or ED, 
in which many-body effects are accounted for \emph{without} breaking the spin symmetry. 
In this case, we find that the SOMO transmission resonance is split 
in \textit{each} spin channel,  
revealing an underlying transmission node that is absent in the symmetry-breaking solution, see Fig.~\ref{fig:Te_pentadienyl}(b). 
% Another possible mechanism to split the SOMO has a many-body origin, 
% which we investigate by taking into account the Coulomb repulsion 
% \emph{without} lifting the spin degeneracy (i.e., within either R-DMFT or ED). 
% We find that the SOMO transmission resonance is split,  
% revealing an underlying transmission node, see Fig.~\ref{fig:Te_pentadienyl}(c). 
Hence, many-body calculations predict a strong suppression of the conductance, 
by several orders of magnitude, in stark contrast with the single-particle picture, 
where electron transport is dominated by a nearly-resonant ballistic channel through the SOMO.  
Note that the splitting is substantially larger in ED than in R-DMFT, 
and considering that the antiresonance is not aligned with $E_F$, 
it also results in a much stronger suppression of the conductance 
$G=1.2 \times 10^{-4} \ G_0$ (ED) versus $G=7.4 \times 10^{-2} \ G_0$ (R-DMFT). 
This suggests that while the driving mechanism has a local nature, 
non-local correlation effects play an important quantitative role, 
as is expected in low-dimensional systems~\cite{valliPRB91,pudleinerPRB99}. 
Such a transmission node in the pentadienyl radical has been observed and analyzed 
within the framework of a single AIM~\cite{bergfieldNL11}. 
Since linear $\pi$-conjugated molecules do not display any topological nodes, 
it has been suggested that this arises from destructive interference 
between different charged states of the molecule~\cite{bergfieldNL11}. 
In Sec.~\ref{sec:mechanism}, we discuss in detail the microscopic mechanism responsible 
for the splitting of the SOMO and for the transmission node, and show that they are intertwined.

\textit{Non-perturbative nature of the splitting } Within ED and R-DMFT, 
%the solution of the many-body problem in the active space 
%(i.e., on the lattice or the auxiliary AIMs) is \emph{numerically exact}. 
%This means that 
the Coulomb repulsion is taken into account in a \emph{non-perturbative} way. 
It is therefore interesting to compare these results to 
a \emph{perturbative} approach, 
such as the GW approximation~\cite{hedinPR139,aryasetiawanRPP61}, 
which has been extensively applied to molecules~\cite{stanEPL76,neatonPRL97,thygesenJCP126,thygesenPRB77,rostgaardPRB81,strangePRB83}. 
This raises the question of how reliably many-body perturbation theory 
can capture the physics of open-shell systems~\cite{mansouriNJP23}. 
The GW approximation arises naturally 
within the context of Hedin equations by neglecting vertex corrections 
(see, e.g., ~\cite{held2011hedin}) 
and the GW self-energy is computed %to the lowest order in perturbation theory, 
as a convolution of the Green's function and the screened Coulomb interaction. 
We compute the GW self-energy projected onto the $A$ subspace 
\begin{equation}
 \mathbf{\Sigma}_{\rm GW}(z) =  i \mathbf{G}_A(z) \mathbf{W}_A(z), 
\end{equation}
with the procedure described in~\cite{gandusSSE199}. 
The self-energy can be calculated either using the bare Green's function 
(G$_0$W$_0$ approximation) 
or by self-consistently dressing the Green's function with $\Sigma_{\rm GW}$. 
In Fig.~\ref{fig:Te_GW} we show that 
neither the G$_0$W$_0$ nor the fully self-consistent GW approximation 
can induce a splitting of the SOMO resonance. 
Instead, both yield only a shift of the resonance above the Fermi energy, which in the junction setup, corresponds to a charge redistribution 
from the molecule to the electrodes, rather than a true oxidation of the radical.

Ultimately, we have shown a set of \textit{different} mechanisms 
for the splitting of the SOMO. 
All result in a suppression of the conductance with respect to 
the single-particle DFT simulation, 
but exhibit very different transport properties. 
This suggests that focusing on the value of $G/G_0$ alone, 
one remains \textit{blind} to the underlying microscopic mechanism. 
Importantly, our analysis demonstrates that the many-body techniques 
that we propose to investigate open-shell molecules 
are not only \emph{sufficient} but also \emph{necessary} 
to capture strong correlation effects, 
whereas common, yet less sophisticated, approaches such as GW or DFT+U 
fall short in describing certain electronic and transport properties 
of organic radicals.

\begin{figure}[tp!]
\includegraphics[width=1.0\linewidth, angle=0]{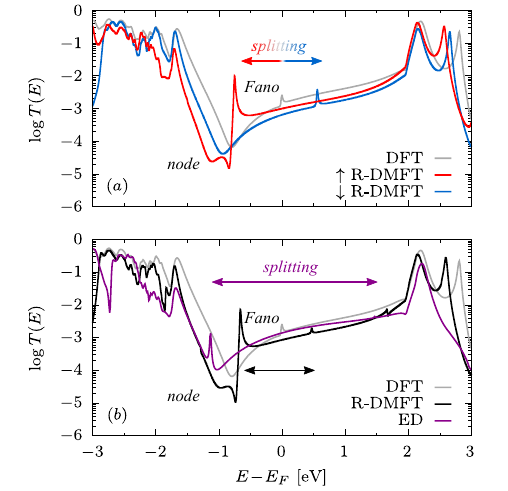}
\caption{Electron transmission function through the benzyl radical junction. 
DFT predicts a SOMO resonance close to $E_F$. Taking into account the Coulomb repulsion beyond restricted DFT yields 
a spin-splitting of the Fano resonance of many-body origin (ED and restricted R-DMFT) or due to the spin-symmetry breaking (unrestricted R-DMFT). 
The QI node present at all levels of theory at $\epsilon_{\mathrm QI}$ arises due to the meta-connection of the substituted benzene ring. }
\label{fig:Te_BDA-CH2}  
\end{figure}

\subsection{Benzyl} 
In single-molecule junctions with a benzene functional unit, 
there are a few possible configurations for the molecule 
bridging the electrodes, 
depending on the relative position of the substituted amino anchoring groups. 
We focus on the \emph{meta} configuration, 
with amino groups at the 1,3-positions of the aromatic ring, 
which exhibits quantum interference (QI) effects 
which are particularly relevant in the context of molecular electronics.

\textit{Single-particle quantum interference.} 
Within DFT, the transmission function displays two striking features 
which can be readily identified in Figs.~\ref{fig:Te_BDA-CH2}(a,b): 
a narrow asymmetric Fano resonance at $\epsilon_{\mathrm{Fano}}<10$~meV, close to $E_F$, 
and a wide antiresonance at $\epsilon_{\mathrm{QI}}\approx -0.8$~eV. 
Both features originate from QI effects.
Clarifying the nature of the resonances and highlighting their differences 
is helpful to understand how electronic correlations affect the transport properties.  
%and to shed light on the underlying microscopic mechanism. 
 
The antiresonance is the hallmark of destructive QI 
%in the meta configuration 
and it is well-established in the literature, 
from both the experimental~\cite{arroyoAC125,yangCCL29,liNatMat18} 
and theoretical~\cite{sautetCPL153,solomonJCP129,samangNJP19,nozakiJPCC121,gunasekaranNL20} points of view. 
% It arises from the interference between the HOMO and LUMO of the ring itself~\cite{gunasekaranNL20}. 
% Such antiresonance is also found in the unfunctionalized benzene junction, 
% but displays a subtle interplay with functional groups (not necessarily radical). 
% It is widely recognized that substituents affect 
% the relative position of destructive interference features with respect to the Fermi energy, 
% see e.g.,~\cite{zhou2021substituent} and references therein. 
% The chemical control of the antiresonance can be exploited for a wide range of applications 
% ranging from nanoelectronics~\cite{zhou2021substituent} to chemical sensing~\cite{prasongkit2016quantum,sengulPRB105enhancing}
It is a common feature in molecular junctions 
with $\pi$-conjugated hydrocarbon bridges. 
%with and without functional groups, 
It arises from the interference between MOs~\cite{gunasekaranNL20} 
and can be exploited for a wide range of applications~\cite{zhou2021substituent} 
including chemical sensing~\cite{prasongkit2016quantum,sengulPRB105enhancing}.

%In principle, the position of the antiresonance is also influenced 
%by the substitution position in the ring (see, e.g.,~\cite{zhou2021substituent} and references therein), 
%but this effect is of marginal relevance to the scope of the present work. 

The resonance with a characteristic asymmetric Fano line shape arises from QI between 
a discrete state and a continuum background.~\cite{fanoPR124,miroshnichenkoRMP82} 
Here, they correspond to the narrow SOMO resonance, 
%which is partially delocalized across the phenyl ring but
which has nodes at the anchoring groups and is thus weakly coupled to the electrodes, 
and the MOs delocalized along the molecular backbone, 
with a strong overlap to the states of the Au electrodes~\cite{zhengAP61}. 
Hence, the Fano resonance can be regarded as the transport signature of the SOMO. 
Due to the different shape of the SOMO resonance in the two radicals, 
i.e., Lorentzian or Breit-Wigner for pentadienyl and asymmetric Fano for benzyl, 
it is interesting to investigate the effect of the Coulomb repulsion 
and highlight the differences between the two cases.  

Within the DFT simulations, the transmission function  displays
a narrow Fano resonance close to $E_F$. 
Due to the presence of the wider QI antiresonance, 
the transmission function is strongly asymmetric across the Fermi energy
and the Fano shape is partially concealed. 

\textit{Symmetry-breaking.}
The spin-unrestricted R-DMFT yields a pair of spin-split Fano resonances, 
as shown in Fig.~\ref{fig:Te_BDA-CH2}(a). 
They originate by the QI interference between 
the $\uparrow$-SOMO and $\downarrow$-SUMO spin-orbitals with the other MOs. 
In the majority spin channel, $\epsilon^{\uparrow}_{\mathrm{Fano}}<0$ 
falls within the energy region in which the transmission is suppressed by the antiresonance, 
and the asymmetric Fano profile is clearly observable. 
Its counterpart in the minority spin channel is found at $\epsilon^{\downarrow}_{\mathrm{Fano}}>0$,   
and is mostly concealed by the background transmission. 
Interestingly, the spin-symmetry breaking also induces 
spin-resolved QI antiresonances, 
already observed in the literature~\cite{valliNL18,valliPRB100,phungPRB102,sengul2021electrode,chelliNL23} 
but the spin splitting $\epsilon^{\downarrow}_{\mathrm{QI}}-\epsilon^{\uparrow}_{\mathrm{QI}}$ 
is weaker than for the Fano resonance. 
This is because the spin imbalance $\langle S_z \rangle \simeq 1/2$ due to the unpaired electron 
is mostly %($\sim 67\%$) 
localized on the p$_z$ LO of the benzylic C atom, 
while the rest of the molecule is weakly magnetic. %($\sim 32\%$) 
Similar results are observed in the literature, 
when electron-electron interactions are included at the static mean-field level within DFT+U 
and the spin symmetry is broken~\cite{smeuJPCC114}. 

\textit{Many-body splitting.} Again, performing many-body simulations without lifting the spin symmetry reveals another scenario, 
as shown in Fig.~\ref{fig:Te_BDA-CH2}(b). 
The difference is twofold. 
We observe a splitting of the Fano resonance in both R-DMFT and ED 
(significantly larger for the latter), 
but no splitting is detected for the QI antiresonance, which is rather shifted away from $E_F$. 
This suggests that the microscopic mechanisms behind the splitting 
with and without spin-symmetry breaking are fundamentally different, 
because they distinguish between the two QI features. 
Moreover, in contrast to the case of pentadienyl, 
the splitting of the SOMO Fano resonance in benzyl does not result 
in a strong suppression of the transmission within the SOMO-SUMO gap. 
The two observations above are deeply connected and  
can be rationalized in terms of the spatial distribution of the SOMO. 
The GW results for benzyl are similar to those for pentadienyl, 
and justify analogous conclusions (not shown).

\section{Microscopic mechanism}\label{sec:mechanism}

\subsection{Splitting of the SOMO}
So far, we have seen that the Coulomb repulsion induces a splitting of the SOMO of the organic radicals. 
In order to gain a deeper understanding of the electronic mechanism behind the splitting, 
and how it affects the transport properties of the junction, 
it is useful to look at the retarded self-energy in the LO basis. 
We adopt the simplified notation $\Sigma_{ij} = \mathbf{\Sigma}_{A;ij}$, 
for both $\mathbf{\Sigma}_A^{\mathrm{ED}}$ and $\mathbf{\Sigma}_A^{\sigma,\mathrm{R-DMFT}}$ 
defined in Eqs.~(\ref{eq:SigmaED}) and (\ref{eq:SigmaDMFT}), respectively. 
The many-body effects encoded in the self-energy can be rationalized by interpreting 
the real part as an energy-dependent level shift,  
and the imaginary part as an effective electron-electron scattering rate. 
We argue that the mechanism discussed in the following 
is a generic feature of organic radicals with a single unpaired electron.  
Therefore, we discuss pentadienyl and benzyl radicals in parallel 
and highlight any differences whenever necessary.

\begin{figure}[bp]
\includegraphics[width=\linewidth, angle=0]{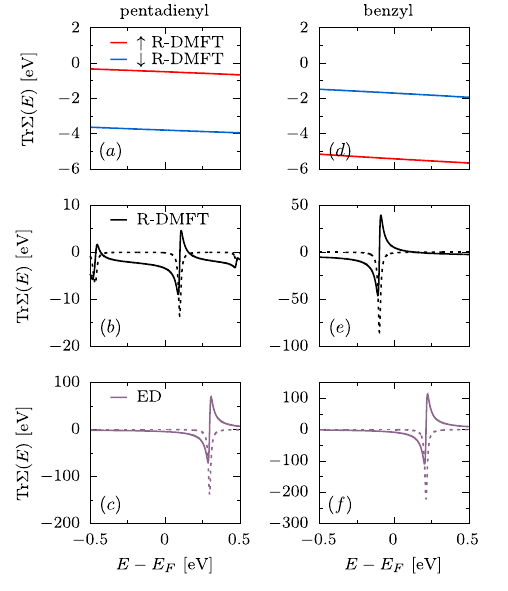}
\caption{Trace of the retarded self-energy $\Tr[\mathbf{\Sigma}(E)]$ in the LO basis 
for the pentadienyl (a, b, c) and benzyl (d, e, f) radicals 
(the real and imaginary parts are denoted by solid and dashed lines, respectively). 
Within spin-unrestricted R-DMFT (a, d) the self-energy displays a weakly energy-dependent real part, 
which is different in each spin sector, while the imaginary part is negligible (not shown). 
Within both R-DMFT (b, e) and ED (c, f) the self-energy is dominated by a single resonance. }
%at energy $\epsilon_r$ (denoted by a solid grey line). }
\label{fig:self_trace}  
\end{figure}

In order to compare the different approximations, it is convenient to look at the trace of the self-energy matrix. 
Within spin-unrestricted R-DMFT, which is shown in Figs.~\ref{fig:self_trace}(a,d), 
the real part of the self-energy is weakly energy-dependent around $E_F$,  
and determines a shift of the SOMO resonance in opposite directions 
for the two spin polarizations. 
The imaginary part is negligible (not shown) 
resulting in highly coherent SOMO and SUMO electronic excitations below and above $E_F$.
%at $\epsilon^{\downarrow}_{\mathrm{SOMO}}$ and $\epsilon^{\uparrow}_{\mathrm{SOMO}}$, 
Note that the ground state of spin-unrestricted R-DMFT is two-fold degenerate, 
and it is invariant under a flip of all spins: $\{\sigma_i\}\rightarrow\{\bar{\sigma}_i\}$. 
This picture is qualitatively analogous to what one can expect 
also at the single-particle level, i.e., within DFT+U. 
Many-body effects are weak, and the dominant effect arises from the spin-symmetry breaking, 
as both radicals are magnetic insulators with a spin SOMO-SUMO gap.

The scenario is completely different within restricted R-DMFT and ED,  
as shown in Figs.~\ref{fig:self_trace}(b,c,e,f).  
There, the self-energy is dominated by a single resonance
and its energy dependence can be well described within a one-pole approximation (OPA)
\begin{equation} \label{eq:selfOPA}
    \Sigma_{\mathrm{OPA}}(E) = \frac{a^2}{E-E_F-\epsilon_{\mathrm{res}} + \imath \gamma}. 
\end{equation}
The OPA self-energy has a Lorentzian shape, where $\epsilon_{\mathrm{res}}$ and $\gamma$ denote 
the resonant energy and the width of the resonance, whereas $a$ controls the amplitude.  
In the simulations, the imaginary part of the self-energy plays the role of a \emph{giant} electron-electron scattering rate 
and suppresses electronic excitations around $\epsilon_{\mathrm{res}} \simeq \epsilon_{\mathrm{SOMO}}$, 
while the real part redistributes the spectral above and below the resonance energy. 
This many-body mechanism, akin to the Mott metal-to-insulator transition as described within DMFT~\cite{georgesRMP68},   
is at the origin of the splitting of the SOMO resonance. 
Remarkably, this feature is completely absent in the GW self-energy, 
whose real part displays a weak energy dependence, 
whereas the imaginary part is negligible close to the SOMO resonance (not shown) 
confirming the non-perturbative nature of the splitting. 

In organic radicals, a hierarchy of emergent energy scales is realized: 
$\Gamma_{\mathrm{SOMO}} \ll \Delta < \Delta_0 \lesssim U_{\mathrm{screened}}$, 
where $\Delta_0 \sim {\rm eV}$ denoted the HOMO-LUMO gap, controlled by the C-C $\pi$-bonds, 
and $\Delta$ is the SUMO-SOMO many-body splitting. 
Hence, the typical energy scale associated with the screened Coulomb repulsion $U_{\mathrm{screened}}$ 
significantly exceeds the width of the SOMO resonance ($\Gamma_{\rm SOMO}\sim 10$--$100$~meV), 
which is controlled by the strength of the coupling to the electrodes. 
This scenario sets the electrons of the radicals considered here 
in the strongly correlated regime. 
Since these are typical energy scales for organic molecules, 
we argue that this mechanism is general for organic radicals with a single unpaired electron. 
In contrast, in closed-shell configurations, electronic correlations 
only result in a many-body renormalization of the single-particle gap~\cite{senterfPRB80,hueserPRB87,valliPRB86,valliPRB91,valliPRB94,schuelerEPJST226,valliNL18,valliPRB100,pudleinerPRB99,fediaiSD1}. 
Multi-radical molecules~\cite{mishra2020topological} and radical networks~\cite{alcon2022unveiling}, 
are expected to display different electronic and transport properties 
due to effective interactions between the unpaired electrons~\cite{mishraNat598,turco2021surface,jacobPRB106,zhengNC11}.

\begin{figure*}[tp]
\includegraphics[width=\linewidth, angle=0]{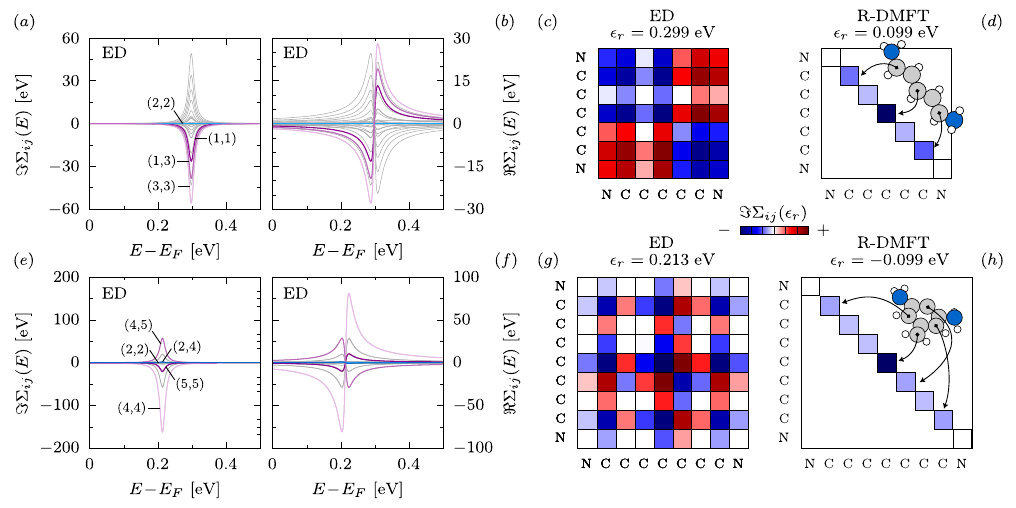}
\caption{Component of the ED self-energy $\Sigma_{ij}(E)$ and its matrix representation 
at the resonant energy $\Im\Sigma_{ij}(\epsilon_{\mathrm{res}})$ in the LO basis 
for the pentadienyl (a,b,c,d) and benzyl (e,f,g,h) radicals. 
Each component of the self-energy (grey lines) is dominated by a single pole (a,b,e,f) at a resonant energy $\epsilon_r$. 
Selected components $(i,j)$ are highlighted (color lines) and are labeled according to their index in the matrix, starting with $\Sigma_{00}$ for the left N-p$_z$ LO. 
The matrix structure of the self-energy reflects the spatial distribution of the SOMO, 
i.e., the largest local ($\Sigma_{ii}$) and non-local ($\Sigma_{ij\neq i}$) self-energy contributions 
are found for the LOs with the largest projections to the SOMO 
(denoted by arrows, see also Fig.~\ref{fig:radicals_SOMO}). 
Within R-DMFT (d,h) the self-energy is diagonal in the LO indices $\Sigma_{ij}\propto\delta_{ij}$ and displays the same pattern. }
\label{fig:self_matrix}  
\end{figure*}

\subsection{Spatial structure of the electronic correlations}\label{sec:locality}
So far, R-DMFT and ED seem to qualitatively describe the same many-body mechanism 
for the splitting of the SOMO. 
However, besides the trace of the self-energy, 
it is interesting to visualize the whole self-energy matrix. 
As discussed in Sec.~\ref{sec:many-body}, 
within ED all elements $\Sigma_{ij}\neq 0$, whereas within R-DMFT $\Sigma_{ij} \propto \delta_{ij}$. 
Remarkably, all elements of the self-energy (regardless of the approximation) 
are well described by the OPA with the \emph{same} resonant energy $\epsilon_r$, 
as shown in Figs.~\ref{fig:self_matrix}(a, e). 
The off-diagonal elements can have either sign, but causality requires $\Im\Sigma_{ii}<0$.    
It is easier to have a comprehensive look at the self-energy 
by plotting the matrix $\Sigma_{ij}(\epsilon_r)$, 
as shown in Figs.~\ref{fig:self_matrix}(c, d, g, h). 
Indeed, looking at the ED self-energy matrix, clear patterns emerge.
Along the diagonal, some elements $\Sigma_{ii}$ are significantly larger than the others 
(note the logarithmic scale), 
and this asymmetry is mirrored by the off-diagonal elements. 
Upon close inspection, we can associate them with the p$_z$ LOs with the largest SOMO projection, 
thus confirming that the strongest many-body effects correlate 
with the spatial distribution of the SOMO. 
Within R-DMFT, we find an analogous pattern along the diagonal, as indicated by the insets molecules. 

Despite its approximations (local Coulomb interaction, local correlations), 
restricted R-DMFT seems to tell qualitatively the same story as the full ED simulations. 
This advocates for a substantially local character of the microscopic mechanism, 
that can describe both the splitting of the SOMO and its consequences on electron transport, 
whereas non-local effects only result in a \textit{quantitative} renormalization of the SOMO-SUMO splitting.

\begin{figure*}[htp]
\includegraphics[width=\linewidth, angle=0]{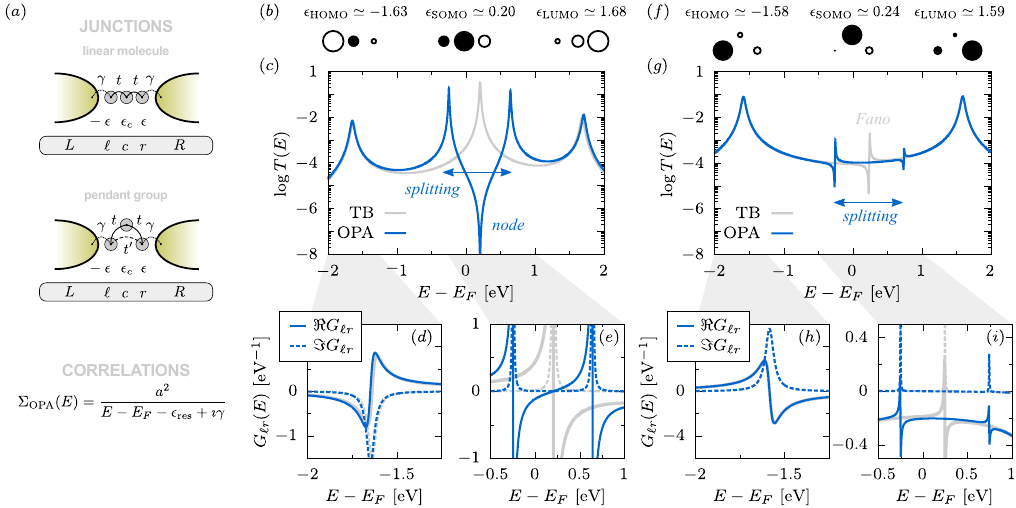}
\caption{(a) Schematic representation of the two junction topology described by 
the three-orbital TB model with its parameter, and form of the OPA self-energy. 
(b,f) Weight distribution and eigenvalues of the TB MOs, in units of [eV].  
Each MO is represented with circles centered at the position of the AOs, 
the circle size reflects the weight of the MO projection on the AOs,  
and the color (black/white) indicates the sign of the coefficient.
(c,g) The transmission function obtained without (grey lines) and with (blue lines) the OPA self-energy. 
The OPA approximation captures all qualitative features of the \textit{ab-initio} + many-body simulations. 
(d,e,h,i) The Green's function $G_{\ell r}$ is shown for specific energy ranges, 
which are relevant to explaining the spectral features associated with the HOMOs (d,h) and the SOMOs (e,i), 
as discussed in the text. 
Model parameters, in units of [eV]: $\epsilon=1.5$, $\epsilon_c=0.25$, $a=0.5$, $\Gamma=0.05$, $\gamma=0.003$, common to both scenarios, 
linear molecules are represented by choosing $t=0.5$, $t'=0$ (b,c,d,e) 
whereas molecules with a pendant group are represented by choosing $t=0.1$, $t'=0.5$ (f,g,h,i). }
\label{fig:model}  
\end{figure*}

% \subsection{Inelastic electron scattering} 
% As anticipated in Sec.~\ref{sec:corrQT}, corrections to the ballistic (Landauer) transmission  
% are expected to be relevant in the presence of inelastic electron-electron scattering 
% due to the Coulomb repulsion. 
% Surprisingly, numerically, we find that vertex corrections are \textit{negligible} 
% for both molecular radicals considered here. 
% This can be understood considering the structure of the vertex correction. 
% In the LO basis, the inelastic scattering rate $\mathbf{\Gamma}_{D}$ is sizable 
% at $E \approx E_{\mathrm{SOMO}}$, around which the many-body self-energy is peaked 
% for the C-p$_z$ LOs and negligible for the (nearly fully-filled) N-p$_z$ LOs, 
% cfr. Figs.~\ref{fig:self_trace} and \ref{fig:self_matrix}. 
% Instead the molecule-electrode coupling matrices $\mathbf{\Gamma}_L$ and $\mathbf{\Gamma}_R$ 
% are large for the N-p$_z$ LOs, which bond with the Au surface. 
% As a result, the vertex $\mathbf{\Lambda}$ is negligibly small. 
% Whether this unexpected property is a generic consequence of 
% weak molecule-electrode bonds mediated by anchoring groups, 
% is an open question that we leave for future investigations. 
% \AV{[This holds in R-DMFT but it does not seem to hold for ED because: 
% (i) non-local terms $\Sigma_{ij}$ between the N-p$_z$ and the C-p$_z$ LOs are non-zero, 
% (ii) in pentadienyl, even $\Sigma_{ii}$ is non-zero for the N-p$_z$ LOs...
%      are the vertex corrections negligible in ED? (verify numerically)]}

\subsection{Implications for electron transport: insights from a minimal model}
The many-body mechanism behind the splitting of the SOMO is common to both the pentadienyl and benzyl radicals.  
However, its consequences on electron transport are dramatically different. 
To understand why, we combine the insights from DFT 
and the knowledge of the spatial and energy structure of the self-energy 
as follows. 

In pentadienyl, the SOMO is delocalized throughout the molecular backbone, 
and its large projection on the p$_z$ LOs of the amino anchoring groups, 
see Fig.~\ref{fig:radicals_SOMO}(a), 
ensures a substantial overlap with the states in the metallic electrodes 
and establishes a transmission channel across the junction through the SOMO. 
However, the pole of the self-energy results in a zero of the Green's function.
%i.e., $G_{\ell r}(\epsilon_r)=0$, with $\ell$ and $r$ the Np$_z$ LOs. 
which hinders electron transport at that energy 
and is at the origin of the transmission node~\cite{valliNL18,valliPRB100}. 
In contrast, in the benzyl radical, the SOMO 
has negligible projection on the amino anchoring groups,
see Fig.\ref{fig:radicals_SOMO}(d), 
and transport is dominated by transmission channels involving the frontier MOs. 
Therefore, the splitting of the Fano resonance weakly affects those channels   
and does not hinder off-resonant transmission of electrons across the junction.

The above picture can be reproduced within a minimal tight-binding (TB) model, 
which is schematically represented in Fig.~\ref{fig:model}(a).  
Let us consider three AOs ($\ell$, $c$, $r$) 
that can be interpreted as the C-p$_z$ orbitals, whereas the anchoring groups are absorbed in an effective description of the reservoirs.  
The Hamiltonian in such a basis reads
\begin{equation}
    \mathbf{H} = \begin{pmatrix}
     \epsilon_{\ell} & t          & t' \\
     t               & \epsilon_c & t  \\
    t'               & t          & \epsilon_{r}
    \end{pmatrix}.
\end{equation}
The hybridization to the electrodes is mediated by the external orbitals ($\ell$, $r$) which bond to the anchoring groups, 
and for the sake of this discussion, it is assumed to be energy-independent: 
\begin{equation}
    \mathbf{\Gamma}_L = \begin{pmatrix}
     \Gamma & 0 & 0 \\
     0      & 0 & 0 \\
     0      & 0 & 0
    \end{pmatrix}, \
    \mathbf{\Gamma}_R = \begin{pmatrix}
     0 & 0 & 0 \\
     0 & 0 & 0 \\
     0 & 0 & \Gamma
    \end{pmatrix}.
\end{equation}
The Hamiltonian of the isolated system can be diagonalized to obtain the eigenvalues 
$\epsilon_{\mathrm{HOMO}}$, $\epsilon_{\mathrm{SOMO}}$, and $\epsilon_{\mathrm{LUMO}}$. 
The Green's function of the device is given by
\begin{equation}
    \mathbf{G}_D(z) = \big[ z - \mathbf{H} 
                              + \imath\mathbf{\Gamma}_L/2 + \imath\mathbf{\Gamma}_R/2 
                              - \mathbf{\Sigma}_D(z) \big]^{-1}
\end{equation}
and is \textit{dressed} with the retarded self-energy of the device $\mathbf{\Sigma}_D(z)$.  
To pinpoint the microscopic mechanism behind the many-body splitting of the SOMO resonance, 
we employ an OPA self-energy of the form 
\begin{equation} 
    \mathbf{\Sigma}_D(z) = \begin{pmatrix}
    0 & 0 & 0 \\
    0 & \Sigma_{\mathrm{OPA}}(z) & 0 \\
    0 & 0 & 0
    \end{pmatrix} 
\end{equation}
with 
\begin{equation}
 \Sigma_{\mathrm{OPA}}(z) = \frac{a^2}{z-\epsilon_{\mathrm{res}}}, 
\end{equation} 
where $z = E-E_F+\imath\gamma$, 
with $a$ controlling the weight of the self-energy, 
and a self-energy pole located at the position of the SOMO resonance $\epsilon_{\mathrm{res}} \simeq \epsilon_{\rm SOMO}$.  
The OPA self-energy  resembles the form of the self-energy obtained from the \emph{ab-initio} + many-body simulations, 
for which the largest contribution is the local element of the radical C atom;
see Figs.~\ref{fig:self_matrix}(a-e) for a comparison. 
%note that in real space, the SOMO is strongly localized at site $c$, 
%and it is well approximated by the corresponding AO i.e., 
%$\langle \psi^{\mathrm{AO}}_c|\psi^{\mathrm{MO}}_{\mathrm{SOMO}}\rangle \approx 1$. 

Within such a three-orbital model, the Landauer transmission in Eq.~(\ref{eq:transmission}) reduces to 
\begin{equation}
    T(E) = \Gamma^2 |G_{\ell r}(E)|^2, 
\end{equation}
where $G_{\ell r}=\mathbf{G}_{D; \ell r}$ is the only element of the Green's function 
that survives the trace in the Landauer formula,  
given the form of the hybridization matrices $\mathbf{\Gamma}_L$ and $\mathbf{\Gamma}_R$,  
and therefore describes the only transmission channel across the junction~\cite{valliJCE22}. 

For simplicity, one can fix $-\epsilon_\ell = \epsilon_r = \epsilon$, and $\epsilon_c \ll \epsilon$, 
together with $a$, $\Gamma$, and $\eta$, 
and choose the parameters $t$ and $t'$ to describe two scenarios, 
which are representative of the pentadienyl and benzyl radicals. 
The results are presented in Fig.~\ref{fig:model} and discussed below. 

The physics of the pentadienyl radical (or in general, a linear conjugated molecule with an \textit{odd} number of C atoms) 
can be reproduced by choosing $t\lesssim\epsilon$ and $t'=0$. 
Hence, electron transport occurs via tunneling processes between adjacent atoms, 
resembling \textit{through bond} couplings between the corresponding AOs with amplitude $t$. 
The corresponding SOMO is fairly delocalized along the molecular backbone, 
with the highest projection of the central AO, cfr. Fig.~\ref{fig:radicals_SOMO}(b) and Fig.~\ref{fig:model}(b). 
The TB transmission function in Fig.~\ref{fig:model}(c) displays a SOMO resonance 
that is split by including the OPA self-energy,  
thus revealing a transmission node within the SOMO-SUMO gap. 
The origin of the transmission node is ascribed to a zero of the Green's function at the SOMO energy 
$G_{\ell r}(E\simeq\epsilon_\mathrm{SOMO}) \approx 0$~\cite{valliNL18,valliPRB100,valliCarbon214},  
as demonstrated in Fig.~\ref{fig:model}(e). 

Instead, with the choice of parameters $t \ll t' \lesssim \epsilon$, one can describe the physics of the benzyl radical, 
characterized by an orbital $c$, which is weakly coupled to the $\ell-r$ molecular backbone. 
The corresponding SOMO is fairly localized on the orbital playing the role of the \textit{pendant} group, 
with little weight on the orbitals connected to the reservoirs,   
cfr. Fig.~\ref{fig:radicals_SOMO}(d) and Fig.~\ref{fig:model}(f). 
The transmission function in Fig.~\ref{fig:model}(g) displays a sharp Fano resonance that is split by the OPA self-energy.  
In contrast to the previous case, $G_{\ell r}$ does not have a zero 
within the SOMO-SUMO gap, 
and transport is dominated by a transmission channel along the molecular backbone, 
corresponding to the direct $\ell$-$r$ hopping process with amplitude $t'$. 
Finally, note that in both scenarios above, many-body effects are negligible for the HOMO and LUMO resonances 
(corresponding to orbitals that are either completely filled or empty) 
irrespective of whether the ``correlated" $c$ orbital has a sizable hybridization with $\ell$ and $r$, 
as can be inferred comparing Figs.~\ref{fig:model}(d,h).  

Such a minimal model can reproduce all fundamental features 
for both radical junctions discussed in this work 
just by changing the parameters of the single-particle Hamiltonian 
to represent one or the other molecule.  
At the same time, it provides a simple and direct interpretation 
of the numerical simulations. 

The splitting mechanism can be better understood by further reducing the minimal model 
to a single level at $\epsilon_c$ renormalized by an OPA self-energy, 
resonant at $\epsilon_{\mathrm{res}}$. 
The corresponding Green's function reads
\begin{equation}
    G(\omega) = \frac{1}{\omega - \epsilon_c - \Sigma_{\mathrm{OPA}}(\omega)},
\end{equation}
with $\Sigma_{\mathrm{OPA}}(\omega)$ given by Eq.~(\ref{eq:selfOPA}). 
It can be shown that such a Green's function can be brought into a sum of two simple poles 
\begin{equation} \label{eq:2PA}
    G(\omega) = \left[ \frac{b_-}{\omega - \omega_- + \imath \eta_-} + 
                       \frac{b_+}{\omega - \omega_+ + \imath \eta_+} \right], 
\end{equation}
where the poles' positions $\omega_{\pm}$, residues $b_{\pm}$, and damping factors $\eta_{\pm}>0$ 
can be calculated exactly (see Appendix~\ref{app:A}). 
In the \textit{strong coupling} limit $a/\gamma \gg 1$ the splitting between the poles 
$\omega_+-\omega_- \approx 2a$ and the damping $\eta_{\pm} \approx \gamma/2$, 
so that the two resonances are spectroscopically resolved. 
%This mechanism is representative of the physical situation for the two organic radicals.  
%Whether the single-particle resonance has a Breit-Wigner or asymmetric Fano shape, 
%depends on the couplings between the $\{\ell, c, r\}$ AOs, as discussed above. 

%In the opposite limit $a \to 0$, 
%the self-energy vanishes $\Sigma_{\mathrm{OPA}} \to 0$, 
%and the Green's function describes a single resonance at $\epsilon_{\rm SOMO}$. 

Importantly, the residues $b_{\pm}$ are in general \textit{complex}, 
and the Green's function does not reduce to the sum of two independent resonances. 
As a consequence, the transmission function $T(\omega) \propto |G(\omega)|^2$, 
with  
%\begin{equation}
%    |G|^2 = |G_{+}|^2 + |G_{-}|^2 + G_{+}G^\dagger_{-} + G^\dagger_+G_-. 
%\end{equation}
\begin{align}
    |G(\omega)|^2 &= \frac{|b_+|^2}{|\omega-\omega_++\imath\eta_+|^2} + \frac{|b_-|^2}{|\omega-\omega_-+\imath\eta_-|^2}  \\
                  &+ 2\Re\left[ \frac{b_+ b^*_-}{(\omega-\omega_++\imath\eta_+)(\omega-\omega_--\imath\eta_-)}\right] . 
\end{align}
The first two contributions correspond to tunneling probabilities 
through the split SUMO and SOMO resonances 
at $\omega_{\pm} \approx (\epsilon_{c}+\epsilon_{\mathrm{res}})/2 \pm a$, 
whereas the mixed contribution is responsible for QI. 
In the specific case we considered, 
the mixed contribution is \textit{negative} for $\omega_- \lesssim \omega \lesssim \omega_+$  
and suppresses the transmission probability at energies between the resonances,  
resulting in a QI node at $\omega = \epsilon_{\mathrm{res}}$ 
(resonant condition for $\Sigma_{\mathrm{OPA}}$). 
When $\epsilon_{c} = \epsilon_{\mathrm{res}}$, the splitting is symmetric around the QI node.

\section{$I-V_b$ characteristics}

\begin{figure*}[htp]
\includegraphics[width=\linewidth, angle=0]{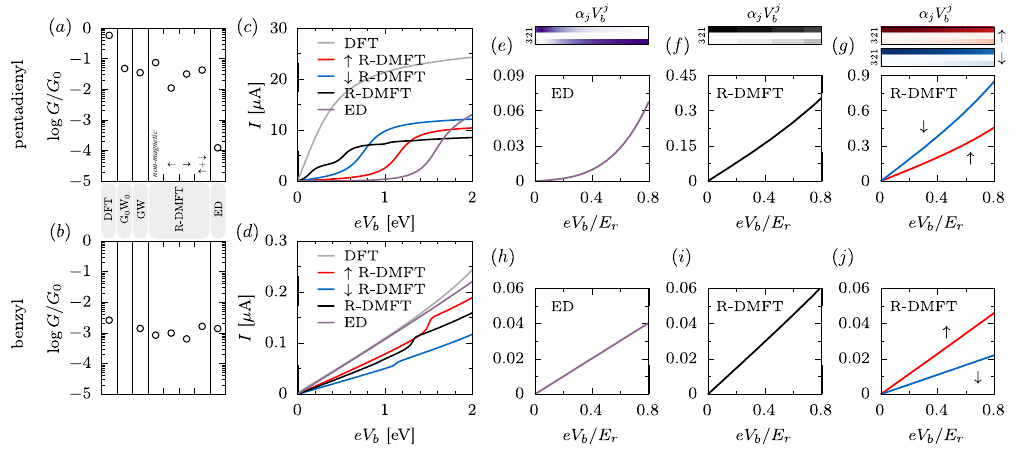}
\caption{(a, b) Comparison of the electron conductances $G/G_0$ and (c, d) current-voltage ($I-V_b$) characteristics 
obtained at different levels of theory, for both the pentadienyl and benzyl molecular junctions. 
(c,d) Comparison of the $I-V_b$ curves from the different theoretical approaches.
(e,f,g) $I-V_b$ characteristics in the low-energy bias window in pentadienyl, 
measured with respect to the lowest transmission resonance at energy $|E_r|$. 
For each value of $V_b$ we perform a numerical fit of the curve with 
$I(V_b) = \sum_{j=1}^{3} \alpha_j V_b^j$, 
and the colorbars above each panel show the relative weight of the $j$-th term. 
Note that the quadratic term is zero because of integrating 
an odd function over a symmetric bias window. 
Non-linear contributions are significant in ED but weaker in R-DMFT (see the text for a detailed discussion). 
(h,i,j) $I-V_b$ characteristics in the low-energy bias window in benzyl, 
measured with respect to the lowest transmission resonance at energy $|E_r|$. 
The narrow Fano resonance does not result in an appreciable deviation from Ohmic behavior. }
\label{fig:GIV}  
\end{figure*}

% old caption
% \caption{(a, b) Comparison of the electron conductances $G/G_0$ and (c, d) current-voltage ($I-V_b$) characteristics 
% obtained at different levels of theory, for both the pentadienyl and benzyl molecular junctions. 
% (c,d) Comparison of the $I-V_b$ curves from the different theoretical approaches.
% (e,f,g) Highlight of the non-linear transport regimes in pentadienyl, 
% indicating the presence/absence of a QI node.  
% Each curve at the corresponding level of theory (grey solid line) 
% is approximated by the expansions $I^{(n)} = \sum_{j=1}^{n} \alpha_j V_b^j$ (color lines). 
% The colorbars above the panels for pentadienyl show the relative weight of the $j$-th term. 
% (h,i) The $I-V_b$ characteristic in benzyl is linear in a wide bias range, 
% barely affected by the narrow Fano resonance(s). See the text for a discussion. }

So far, we have demonstrated that different microscopic mechanisms
underlying the splitting of the SOMO resonance
(i.e., the many-body or magnetic splitting of the SOMO resonance), 
a Fermi energy alignment, or the oxidation of the radical, 
result in different distinctive features in the electron transmission function. 
A natural question is whether it is possible to discriminate between these scenarios. 
Experiments have direct access to the current-bias $I-V_b$ characteristics, 
but a possible strategy to probe the transmission function is to consider 
the differential conductance in the low-bias regime, where $dI/dV_b(V_b)$ resembles $T(E)$. 
It has been shown~\cite{guedonNatNanotech7} that in the presence of an antiresonance at $\omega_{\rm QI}$ 
a ``\textit{v-shaped}'' differential conductance is measured at low bias, 
whereas a ``\textit{u-shaped}'' differential conductance typically suggests constructive QI instead.  
However, the profile of the differential conductance is sensitive to the relative alignment of $\omega_{\rm QI}$ and the Fermi energy, 
and is not entirely conclusive~\cite{guedonNatNanotech7}. 

An alternative fingerprint of destructive QI is a strongly non-linear $I-V_b$ characteristics 
in the non-resonant transport regime~\cite{greenwaldNatNano3,valliCarbon214}. 
In Fig.~\ref{fig:GIV}, we show that, in the case of pentadienyl, 
a QI node in the transmission function corresponds to a strong non-linearity $I-V_b$ curve 
that is already dominant before the onset of the resonant transport regime. 
The underlying argument is simple. 
The current in terms of the transmission function is given by 
\begin{equation}
    I = \frac{e}{h} \int_{-\infty}^{\infty} {\rm d}\omega \, T(\omega) 
                         \left[ f_L(\omega) - f_R(\omega)\right],
\end{equation}
where $f_L(\omega) = f(\omega+eV_b/2)$ and $f_R(\omega)=f(\omega-eV_b/2)$, 
are the Fermi-Dirac distributions of the left and right electrodes, respectively, 
with $f(\omega)=(1+e^{-\omega/k_BT})^{-1}$. 
Near equilibrium, i.e., $T(\omega, eV_b) \approx T(\omega)$, 
off-resonance, and in the limit $k_BT \ll eV_b$, two scenarios are possible: 
\begin{itemize}
 \item[(i)] in the absence of a QI node, the transmission function is nearly constant $T(\omega \approx 0) \approx \tau_0$, yielding a linear characteristics, $I(V_b) \propto V_b$;
 \item[(ii)] in the presence of a QI node at $\omega_{\mathrm{QI}}$, it can be shown that 
 $T(\omega \to \omega_{\mathrm{QI}}) \approx \tau_0 + \tau_2 \, (\omega-\omega_{\rm QI})^2$, 
thus yielding  both a linear and a cubic contribution upon integration: $I(V_b) \approx \alpha_1 V_b + \alpha_3 V_b^3$. 
\end{itemize}
The behavior of the transmission function follows from the analysis of the real-space retarded Green's function, 
for which the real part linearly change sign across the QI node, 
$\Re G_{\ell r}(\omega \to \omega_{\mathrm{QI}}) \propto \omega -\omega_{\rm QI}$~\cite{valliNL18,valliPRB100,valliCarbon214}.

It is useful to analyze the coefficients and, in particular, the effective curvature 
$r = (\alpha_3/\alpha_1) (eV_b)^2$ given by the ratio of the linear and cubic terms 
(note that the quadratic term is zero because of integrating 
an odd function over a symmetric bias window). 
Considering the minimal model in the regime in which the splitting 
is spectroscopically resolved ($a\gamma \gg 1$) 
one typically expects a linear regime ($r<1$) at sufficiently low bias, 
and an \textit{extended} non-linear regime ($r>1$) far from resonant tunneling, 
which, in terms of the parameter of our model, is roughly equivalent to 
$eV_b/2 \lesssim a-|\omega_{\mathrm{QI}}|$. 
Close to resonant transport, additional non-linear terms arise 
because the transmission has non-negligible contributions from the resonance. 
Therefore, far from resonant tunneling, a strong non-linear $I-V_b$ characteristic 
can be interpreted as a fingerprint of destructive QI. 
These considerations provide a clear interpretation of the numerical results. 

%In the limit of \textit{narrow resonances} $\Gamma \ll \epsilon$ 
%and for $\omega_{\rm SOMO} \ll \epsilon$, 
%the linear coefficient $\alpha_1 \propto \Gamma^4$ and 
%the linear contribution dominates ($r < 1$) at bias $eV_b \lesssim \Gamma$, 
%whereas the non-linear (cubic) term dominates ($r > 1$) in the regime $\Gamma < eV_b \ll \epsilon$. 

In Fig.~\ref{fig:GIV} we compare the conductance $G/G_0$ and the $I-V_b$ characteristics 
obtained for both radical junctions at each level of theory. 
For pentadienyl, see Fig.~\ref{fig:GIV}(a), the first observation is that the conductance alone is not sufficient to discriminate between the different scenarios. 
Apart from the DFT+NEGF transport simulations, 
all simulations that include many-body effects yield a conductance 
that is incompatible with nearly-resonant transport. 
In particular, ED yields the lowest conductance because of the wide gap and deep QI node. 
In the spin-restricted R-DMFT, the conductance is deceptively higher 
than in the spin-unrestricted case ($G > G^{\downarrow} > G^{\uparrow}$) 
because of the narrow SOMO-SUMO splitting. 
This is also reflected in the $I-V_b$ characteristics. 
In Fig.~\ref{fig:GIV}(c) we show the current as a function of bias 
obtained at all levels of theory. 
The rapid increase of the DFT current at low bias 
is consistent with the nearly-resonant transport regime.  

In Figs.~\ref{fig:GIV}(e,~f,~g) instead, we highlight the different behavior 
of the many-body simulations by rescaling the bias to $E_r$, 
corresponding to the distance of the closest transport resonance from the Fermi energy. 
ED displays a low current and a strongly non-linear characteristic, 
with the cubic term $\alpha_3V^3$ dominant already at $eV_b/E_r \gtrsim 0.2$. 
This is the hallmark of the QI node within the wide SOMO-SUMO gap. 
In contrast, R-DMFT displays a nearly linear characteristic. 
The rise of a non-linear term at $eV/E_r \gtrsim 0.4$ can be ascribed to the QI node, 
but it is weak because (i) the SOMO-SUMO splitting is narrow and 
(ii) the Fermi level lies very close to one of the split-SOMO resonances. 
This is the worst-case scenario, in which the QI node is present but very difficult to identify 
from the non-linear characteristic. 
In the spin-unrestricted R-DMFT the QI node is absent, 
a weak non-linear term appears in both spin channels only at $eV_b/E_r \gtrsim 0.8$,  
and it is completely associated with the vicinity to the transport resonance at $E_r$. 

For benzyl, we obtain conductance values $G\simeq 10^{-3}G_0$ 
at all levels of theory, see Fig.~\ref{fig:GIV}(b) 
and linear $I-V_b$ characteristic up to $eV_b/E_r \lesssim 1$, 
see Figs.~\ref{fig:GIV}(d,~h,~i,~j). 
This is explained considering that the narrow Fano resonance(s) 
do not generate a contribution to the current large enough to be detected 
over the linear background within the HOMO-LUMO gap.  
So, in this case, the $I-V_b$ characteristic is not useful in identifying the SOMO splitting.

\section{Conclusions}
%In this work, we have proposed a completely \emph{ab-initio} 
%state-of-the-art many-body numerical method that is
%able to address the complexity of a realistic chemical environment 
%as well as electronic correlation effects beyond
%the single-particle picture within a first-principle framework.
In this work, we have unraveled the emergence of strongly correlated electron physics in
radical molecules with a single unpaired electron. 
We identify the splitting of the SOMO resonance as a hallmark of many-body effects, 
with observable consequences in the electron transport in a two-terminal junction setup. 
We argue that this phenomenon can explain puzzling evidence 
such as the low conductance observed in transport measurements.  

Specifically, we performed thorough numerical simulations combining state-of-the-art 
\emph{ab-initio} and many-body techniques. 
Our framework can describe the complexity of a realistic chemical environment  
as well as electronic correlation effects beyond the single-particle picture, 
and it is specifically tailored to address the electron transport properties of quantum junctions. 
By considering a linear and a cyclic radical species, 
we obtain a general understanding of the role of many-body effects 
in molecular radicals with a single unpaired electron,  
and their dramatic consequences on electron transport.  
We establish the microscopic mechanism behind the splitting of the SOMO resonance 
and discover a clear link between the structure of the electronic self-energy 
and the spatial distribution of the SOMO. 
We further demonstrate this through a minimal model 
that reproduces the key phenomenological features of the electron transmission. 
Moreover, we demonstrate that a non-perturbative treatment of the Coulomb repulsion 
is paramount to describe the splitting of the SOMO resonance, 
and alternative common approaches, 
such as many-body perturbation theory or static mean-field (which breaks the spin symmetry) 
fail to reproduce it, resulting in \textit{qualitatively} different features  
and suggesting different interpretations of the underlying physics. 

All approaches that treat the Coulomb repulsion beyond DFT exhibit a suppression of the conductance, 
which is incompatible with nearly-resonant transport through an open channel. 
Beyond the zero-bias conductance, the $I-V_b$ characteristics 
offer an additional perspective on the role of many-body effects. 
In the pentadienyl junction, the correlation-induced splitting of the SOMO gives rise to a spectroscopically resolved transmission node, 
leading to strongly nonlinear $I-V_b$ behavior at low bias. 
This provides an experimentally accessible signature of strong correlations 
by measuring destructive quantum interference that can arise 
neither in the single-particle picture nor in mean-field or perturbative approaches.  
In contrast, for the benzyl junction, the $I-V_b$ curves are qualitatively identical 
across DFT and many-body approaches, reflecting the weak coupling of the SOMO to the electrodes. 
Many-body effects in benzyl remain spectroscopically silent in transport, 
and the narrow Fano resonances are hardly observable in the $I-V_b$ response. 
This underscores the importance of molecular orbital alignment and spatial distribution of the MOs 
in determining whether correlations manifest in measurable transport quantities. 

Finally, we note that the present work neglects vertex corrections in the transmission function, 
which take into account inelastic electron-electron scattering processes. 
%which originate from the dynamical renormalization of the molecule–lead coupling due to the many-body self-energy. 
In open-shell molecular junctions, such as those investigated here, 
these corrections are expected to be sizable, or even the dominant contribution~\cite{valliJCE22}, in the resonant tunneling regime. 
Their inclusion is anticipated to introduce additional broadening and asymmetric reshaping of the transmission and interference features, % particularly around the Fano resonances and interference nodes, 
while preserving the overall qualitative behavior discussed in this work. 
A quantitative evaluation of vertex corrections will be the subject of future investigations.

Overall, our study establishes a framework to understand how strong electronic correlations 
manifest in the transport properties of organic radicals. 
In particular, it highlights that different treatments of the spin degree of freedom 
result in qualitatively different transport properties, 
calling for experiments to discriminate between the different scenarios. 
Thus, our work paves the path toward a deeper and more comprehensive 
understanding of strongly correlated electron physics of organic radicals. \\

\section*{Acknowledgements}
We thank J.~M.~Tomczak for valuable discussions and constructive feedback on the manuscript, 
This research is supported by the Austrian Science Fund (FWF) through project P~31631, 
the HUN-REN Hungarian Research Network through the Supported Research Groups Programme, 
HUN-REN-BME-BCE Quantum Technology Research Group (TKCS-2024/34). 
and the NCCR MARVEL funded by the Swiss National Science Foundation grant 51NF40-205602. 
Computational support from the Swiss Supercomputing Center (CSCS) under project ID s1119 is gratefully acknowledged. \\

\appendix
\section{Two-poles decomposition}\label{app:A}
Let us show how the Green's function
\begin{equation}
   G(\omega) = \frac{1}{\omega-\epsilon_0 - \Sigma_{\mathrm{OPA}}}
\end{equation}
with the one-pole approximation (OPA) self-energy
\begin{equation}
   \Sigma_{\mathrm{OPA}} = \frac{a^2}{\omega-\epsilon_{\mathrm{res}}+i\gamma}.
\end{equation}
can be brought into the sum of two simple poles
\begin{equation} \label{eq:2PA_appendix}
    G(\omega) = \left[ \frac{b_-}{\omega - \omega_- + \imath \eta_-} + 
                       \frac{b_+}{\omega - \omega_+ + \imath \eta_+} \right], 
\end{equation}
where the poles' positions $\omega_{\pm}$, residues $b_{\pm}$, and damping factors $\eta_{\pm}>0$. \\
Let us define $x=\omega-\epsilon_0$ and $\delta = \epsilon_0 - \epsilon_{\mathrm{res}}$. 
Then
\begin{equation}
   G(x) = \frac{1}{x-\dfrac{a^2}{x+\delta+ i\gamma}} = \frac{x+\delta+i\gamma}{x^2 + (\delta+i\gamma) x - a^2}  
\end{equation}
takes the form $G(x) = P(x)/Q(x)$ and the poles are the roots of the quadratic denominator 
\begin{equation}
   x^2 + (\delta+i\gamma) x - a^2 = 0.    
\end{equation}
Let us further define 
\begin{equation}
   \Delta \equiv \sqrt{4a^2 + (\delta + \gamma)^2} %\quad(\text{principal branch})
\end{equation}
then the two roots given by
\begin{equation}
   x_{\pm}=\frac{-(\delta+i\gamma) \pm \Delta}{2}.
\end{equation}
The residues are given by
\begin{equation}
   b_+ = \frac{P(x_+)}{x_+ - x_-}, \qquad  b_- = \frac{P(x_-)}{x_- - x_+}
\end{equation}
with $P(x) = x + \delta +i\gamma$ and $x_+-x_-=\Delta$, we obtain compact forms
\begin{equation}
   b_+ = \frac{x_+ + i\gamma}{\Delta}, \qquad  b_- = -\frac{x_- + i\gamma}{\Delta}.
\end{equation}
Equivalently, substituting the expression found for $x_{\pm}$
\begin{equation}
   b_+ = \frac{1}{2} + \frac{\delta+i\gamma}{2\Delta},\qquad
   b_- = \frac{1}{2} - \frac{\delta+i\gamma}{2\Delta}.
\end{equation}
and one can verify that $b_+ + b_- = 1$, which is consistent with the asymptotic behavior of the Green's function $G(\omega)\sim 1/(\omega-\epsilon_0)$ for $\omega \to \infty$. 

% \section{Taylor expansion of the model transmission function}\label{app:B}
% Given the two-pole decomposition of the Green's function Eq.~(\ref{eq:2PA_appendix}), 
% the corresponding transmission function

\bibliographystyle{apsrev}
\bibliography{bibliography}

\end{document}